\pdfoutput=1
\documentclass[12pt,a4paper]{article}
\usepackage{ifthen} %

\newboolean{pdflatex}
\setboolean{pdflatex}{true} %

\newboolean{articletitles}
\setboolean{articletitles}{true} %

\newboolean{uprightparticles}
\setboolean{uprightparticles}{false} %

\def\paperauthors{LHCb collaboration} %
\def\paperasciititle{Template for writing LHCb papers} %
\def\papertitle{Search for resonance-enhanced $C\!P$ and angular asymmetries in the $\mathit{\Lambda}^+_c\to p\mu^+\mu^-$ decay at LHCb} %
\def\paperkeywords{{High Energy Physics}, {LHCb}} %
\def\papercopyright{\the\year\ CERN for the benefit of the LHCb collaboration} %
\def\paperlicence{CC BY 4.0 licence}
\def\paperlicenceurl{https://creativecommons.org/licenses/by/4.0/}

\newif\ifEnableSectionTOCLinks
\EnableSectionTOCLinksfalse %

\usepackage[top=1in, bottom=1.25in, left=1in, right=1in]{geometry}

\usepackage{microtype}
\usepackage{lineno}  %
\usepackage{xspace} %
\usepackage{caption} %

\usepackage{graphicx}  %
\usepackage{color}
\usepackage{colortbl}
\graphicspath{{./figs/}} %
\usepackage{amsmath} %
\usepackage{amssymb}
\usepackage{amsfonts}
\usepackage{upgreek} %

\newcommand*\patchAmsMathEnvironmentForLineno[1]{%
\expandafter\let\csname old#1\expandafter\endcsname\csname #1\endcsname
\expandafter\let\csname oldend#1\expandafter\endcsname\csname
end#1\endcsname
 \renewenvironment{#1}%
   {\linenomath\csname old#1\endcsname}%
   {\csname oldend#1\endcsname\endlinenomath}%
}
\newcommand*\patchBothAmsMathEnvironmentsForLineno[1]{%
  \patchAmsMathEnvironmentForLineno{#1}%
  \patchAmsMathEnvironmentForLineno{#1*}%
}
\AtBeginDocument{%
\patchBothAmsMathEnvironmentsForLineno{equation}%
\patchBothAmsMathEnvironmentsForLineno{align}%
\patchBothAmsMathEnvironmentsForLineno{flalign}%
\patchBothAmsMathEnvironmentsForLineno{alignat}%
\patchBothAmsMathEnvironmentsForLineno{gather}%
\patchBothAmsMathEnvironmentsForLineno{multline}%
\patchBothAmsMathEnvironmentsForLineno{eqnarray}%
}

\usepackage{hyperref}
\usepackage{hyperxmp}
\hypersetup{pdftex,
pdfauthor={\paperauthors},
pdftitle={\paperasciititle},
pdfkeywords={\paperkeywords},
pdfcopyright={Copyright (C) \papercopyright},
pdflicenseurl={\paperlicenceurl}}

\usepackage[colorinlistoftodos,textsize=scriptsize]{todonotes}

\usepackage[bottom,flushmargin,hang,multiple]{footmisc}

\usepackage[all]{hypcap} %

\usepackage{xspace} 
\usepackage{upgreek}

\newcommand{\Lcpmm}{\mbox{\ensuremath{\Lc\to\proton\mup\mun}}\xspace}
\newcommand{\Lcppp}{\mbox{\ensuremath{\Lc\to\proton\pip\pim}}\xspace}

\def\phil   {{\ensuremath{\phi_{\text{low}}}}\xspace}
\def\phih   {{\ensuremath{\phi_{\text{high}}}}\xspace}

\def\Acp   {{\ensuremath{A_{\CP}}}\xspace}
\def\Acpraw   {{\ensuremath{A_\text{raw}}}\xspace}
\def\AFBcp   {{\ensuremath{A^{\text{\CP}}_{\text{FB}}}}\xspace}

\def\Adet   {{\ensuremath{A_{\text{det}}}}\xspace}
\def\Aprod   {{\ensuremath{A_{\text{prod}}}}\xspace}
\def\AFBLc   {{\ensuremath{A^{\text{\Lc}}_{\text{FB}}}}\xspace}
\def\AFBLcbar   {{\ensuremath{A^{\text{\Lcbar}}_{\text{FB}}}}\xspace}
\def\SAFB   {{\ensuremath{\Sigma A_{\text{FB}}}}\xspace}
\def\DAFB   {{\ensuremath{\Delta A_{\text{FB}}}}\xspace}

\def\mLc {{\ensuremath{m(\proton\mup\mun)}}\xspace}
\def\mmumu {{\ensuremath{m(\mup\mun)}}\xspace}

\newcommand{\integratedresults}[1]{
\begin{align#1}
\Acp &= (-1.1 \pm 4.0 \pm 0.5)\%,\\
\SAFB &= (\phantom{-}3.9 \pm 4.0  \pm 0.6)\%,\\
\DAFB &= (\phantom{-}3.1 \pm 4.0  \pm 0.4)\%,
\end{align#1}
}

\def\lhcb   {\mbox{LHCb}\xspace}

\def\MagUp {\mbox{\em Mag\kern -0.05em Up}\xspace}

\ifthenelse{\boolean{uprightparticles}}%
{

 \def\Pmu         {\ensuremath{\upmu}\xspace}

 \def\Ppi         {\ensuremath{\uppi}\xspace}                 
                  
 \def\Prho        {\ensuremath{\uprho}\xspace}

 \def\Pphi        {\ensuremath{\upphi}\xspace}

 \def\Pomega      {\ensuremath{\upomega}\xspace}                 

 \def\PDelta      {\ensuremath{\Delta}\xspace}                 
 \def\PXi         {\ensuremath{\Xi}\xspace}                 
 \def\PLambda     {\ensuremath{\Lambda}\xspace}                 
 \def\PSigma      {\ensuremath{\Sigma}\xspace}                 
 \def\POmega      {\ensuremath{\Omega}\xspace}                 
 \def\PUpsilon    {\ensuremath{\Upsilon}\xspace}
 \let\oldPi\Pi
 \def\PPi         {\ensuremath{\oldPi}\xspace}

 \def\PB      {\ensuremath{\mathrm{B}}\xspace}                 
 \def\PD      {\ensuremath{\mathrm{D}}\xspace}                 

 \def\PK      {\ensuremath{\mathrm{K}}\xspace}                 
 \def\Pb      {\ensuremath{\mathrm{b}}\xspace}                 
 \def\Pc      {\ensuremath{\mathrm{c}}\xspace}

 \def\Pp      {\ensuremath{\mathrm{p}}\xspace}                 

 \def\Ps      {\ensuremath{\mathrm{s}}\xspace}                 
                  
 \def\Pu      {\ensuremath{\mathrm{u}}\xspace}                 
 \def\thebaroffset{0.0em}
}
{

 \def\Pmu         {\ensuremath{\mu}\xspace}

 \def\Ppi         {\ensuremath{\pi}\xspace}                 
                  
 \def\Prho        {\ensuremath{\rho}\xspace}

 \def\Pphi        {\ensuremath{\phi}\xspace}

 \def\Pomega      {\ensuremath{\omega}\xspace}                 
 \mathchardef\PDelta="7101
 \mathchardef\PXi="7104
 \mathchardef\PLambda="7103
 \mathchardef\PSigma="7106
 \mathchardef\POmega="710A
 \mathchardef\PUpsilon="7107
 \mathchardef\PPi="7105
 \def\PB      {\ensuremath{B}\xspace}                 
 \def\PD      {\ensuremath{D}\xspace}                 

 \def\PK      {\ensuremath{K}\xspace}                 
 \def\Pb      {\ensuremath{b}\xspace}                 
 \def\Pc      {\ensuremath{c}\xspace}

 \def\Pp      {\ensuremath{p}\xspace}                 

 \def\Ps      {\ensuremath{s}\xspace}                 
                  
 \def\Pu      {\ensuremath{u}\xspace}                 
 \def\thebaroffset{0.18em}
}
\newcommand{\offsetoverline}[2][\thebaroffset]{\kern #1\overline{\kern -#1 #2}}%

\makeatletter
\ifcase \@ptsize \relax%
  \newcommand{\miniscule}{\@setfontsize\miniscule{4}{5}}%
\or%
  \newcommand{\miniscule}{\@setfontsize\miniscule{5}{6}}%
\or%
  \newcommand{\miniscule}{\@setfontsize\miniscule{5}{6}}%
\fi
\makeatother

\DeclareRobustCommand{\optbar}[1]{\shortstack{{\miniscule (\rule[.5ex]{1.25em}{.18mm})}
  \\ [-.7ex] $#1$}}

\def\mup        {{\ensuremath{\Pmu^+}}\xspace}
\def\mun        {{\ensuremath{\Pmu^-}}\xspace} %

\def\mumu       {{\ensuremath{\Pmu^+\Pmu^-}}\xspace}

\def\uquark    {{\ensuremath{\Pu}}\xspace}

\def\squark    {{\ensuremath{\Ps}}\xspace}

\def\cquark    {{\ensuremath{\Pc}}\xspace}

\def\bquark    {{\ensuremath{\Pb}}\xspace}

\def\pion   {{\ensuremath{\Ppi}}\xspace}

\def\pip    {{\ensuremath{\pion^+}}\xspace}
\def\pim    {{\ensuremath{\pion^-}}\xspace}

\def\rhomeson {{\ensuremath{\Prho}}\xspace}
\def\rhoz     {{\ensuremath{\rhomeson^0}}\xspace}

\def\kaon    {{\ensuremath{\PK}}\xspace}
\def\Kbar    {{\ensuremath{\offsetoverline{\PK}}}\xspace}

\def\KorKbar {\kern \thebaroffset\optbar{\kern -\thebaroffset \PK}{}\xspace}
\def\Kz      {{\ensuremath{\kaon^0}}\xspace}
\def\Kzb     {{\ensuremath{\Kbar{}^0}}\xspace}
\def\Kp      {{\ensuremath{\kaon^+}}\xspace}
\def\Km      {{\ensuremath{\kaon^-}}\xspace}

\def\KS      {{\ensuremath{\kaon^0_{\mathrm{S}}}}\xspace}

\newcommand{\phiz}{\ensuremath{\Pphi}\xspace}
\newcommand{\omegaz}{\ensuremath{\Pomega}\xspace}

\def\D       {{\ensuremath{\PD}}\xspace}

\def\DorDbar {\kern \thebaroffset\optbar{\kern -\thebaroffset \PD}\xspace}
\def\Dz      {{\ensuremath{\D^0}}\xspace}

\def\Dp      {{\ensuremath{\D^+}}\xspace}
\def\Dm      {{\ensuremath{\D^-}}\xspace}

\def\DpDm    {\ensuremath{\Dp {\kern -0.16em \Dm}}\xspace}

\def\B       {{\ensuremath{\PB}}\xspace}

\def\BorBbar {\kern \thebaroffset\optbar{\kern -\thebaroffset \PB}\xspace}

\def\Bd      {{\ensuremath{\B^0}}\xspace}

\def\BdorBdbar {\kern \thebaroffset\optbar{\kern -\thebaroffset \Bd}\xspace}

\def\Bs      {{\ensuremath{\B^0_\squark}}\xspace}

\def\BsorBsbar {\kern \thebaroffset\optbar{\kern -\thebaroffset \Bs}\xspace}

\def\Y#1S{\ensuremath{\PUpsilon{(#1S)}}\xspace}

\def\proton      {{\ensuremath{\Pp}}\xspace}
\def\antiproton  {{\ensuremath{\overline \proton}}\xspace}

\def\Lz          {{\ensuremath{\PLambda}}\xspace}
\def\Lbar        {{\ensuremath{\offsetoverline{\PLambda}}}\xspace}
\def\LorLbar     {\kern \thebaroffset\optbar{\kern -\thebaroffset \PLambda}\xspace}

\def\Lc          {{\ensuremath{\Lz^+_\cquark}}\xspace}
\def\Lcbar       {{\ensuremath{\Lbar{}^-_\cquark}}\xspace}

\def\BF         {{\ensuremath{\mathcal{B}}}\xspace}

\newcommand{\decay}[2]{\mbox{\ensuremath{#1\!\to #2}}\xspace} 

\def\to                 {\ensuremath{\rightarrow}\xspace}

\def\CP                {{\ensuremath{C\!P}}\xspace}

\def\AFB      {\ensuremath{A_{\mathrm{FB}}}\xspace}

\def\AT#1     {\ensuremath{A_{\mathrm{T}}^{#1}}\xspace}           %

\def\C#1      {\ensuremath{\mathcal{C}_{#1}}\xspace}                       %
\def\Cp#1     {\ensuremath{\mathcal{C}_{#1}^{'}}\xspace}                    %
\def\Ceff#1   {\ensuremath{\mathcal{C}_{#1}^{\mathrm{(eff)}}}\xspace}        %
\def\Cpeff#1  {\ensuremath{\mathcal{C}_{#1}^{'\mathrm{(eff)}}}\xspace}       %
\def\Ope#1    {\ensuremath{\mathcal{O}_{#1}}\xspace}                       %
\def\Opep#1   {\ensuremath{\mathcal{O}_{#1}^{'}}\xspace}                    %

\newcommand{\aunit}[1]{\ensuremath{\text{\,#1}}}       
\newcommand{\tev}{\aunit{Te\kern -0.1em V}\xspace}
\newcommand{\gev}{\aunit{Ge\kern -0.1em V}\xspace}
\newcommand{\mev}{\aunit{Me\kern -0.1em V}\xspace}
\newcommand{\kev}{\aunit{ke\kern -0.1em V}\xspace}
\newcommand{\ev}{\aunit{e\kern -0.1em V}\xspace}
 
\newcommand{\mevc}{\ensuremath{\aunit{Me\kern -0.1em V\!/}c}\xspace}
\newcommand{\gevc}{\ensuremath{\aunit{Ge\kern -0.1em V\!/}c}\xspace}
\newcommand{\mevcc}{\ensuremath{\aunit{Me\kern -0.1em V\!/}c^2}\xspace}
\newcommand{\gevcc}{\ensuremath{\aunit{Ge\kern -0.1em V\!/}c^2}\xspace}

\def\fb   {\ensuremath{\aunit{fb}}\xspace}
\def\invfb   {\ensuremath{\fb^{-1}}\xspace}

\def\gsim{{~\raise.15em\hbox{$>$}\kern-.85em
          \lower.35em\hbox{$\sim$}~}\xspace}
\def\lsim{{~\raise.15em\hbox{$<$}\kern-.85em
          \lower.35em\hbox{$\sim$}~}\xspace}

\def\pt         {\ensuremath{p_{\mathrm{T}}}\xspace}

\def\evtgen     {\mbox{\textsc{EvtGen}}\xspace}

\def\geant      {\mbox{\textsc{Geant4}}\xspace}

\def\photos     {\mbox{\textsc{Photos}}\xspace}

\def\pythia     {\mbox{\textsc{Pythia}}\xspace}

\def\tell1  {TELL1\xspace}
\def\ukl1   {UKL1\xspace}

\newcommand{\lhcborcid}[1]{\href{https://orcid.org/#1}{\hspace*{0.1em}\raisebox{-0.45ex}{\includegraphics[width=1em]{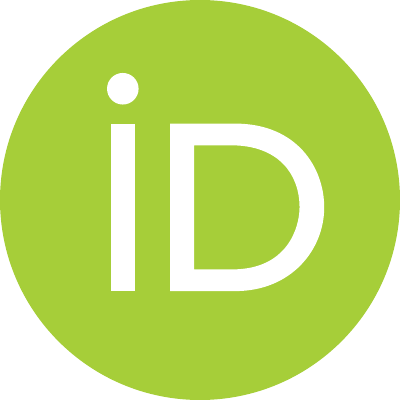}}}}

\hypersetup{
  colorlinks   = true, %
  urlcolor     = blue, %
  linkcolor    = blue, %
  citecolor    = red   %
}

\ifEnableSectionTOCLinks
    \usepackage[explicit]{titlesec} %

    \let\oldcontentsline\contentsline
    \renewcommand

    \titleformat{\section}{\normalfont\Large\bf}{\hyperlink{tocsection.\thesection}{{\thesection} \parbox[t]{\dimexpr\textwidth-1pc}{#1}}}{1pc}{}

    \titleformat{\subsection}{\normalfont\bf}{\hyperlink{tocsubsection.\thesubsection}{{\thesubsection} \parbox[t]{\dimexpr\textwidth-1pc}{#1}}}{1pc}{}

    \titleformat{name=\section,numberless}[display]{}{}{0pt}{\normalfont\Huge\bfseries #1}
\fi

\usepackage{cite} %
\usepackage{mciteplus}
 \usepackage{longtable} %

\newboolean{wordcount}
\setboolean{wordcount}{false} %
\usepackage{booktabs}

\begin{document}

\renewcommand{\thefootnote}{\fnsymbol{footnote}}
\setcounter{footnote}{1}

%
%
%
%
%
%
%
%
%
%
%
%

%
%
%
\begin{titlepage}
\pagenumbering{roman}

\vspace*{-1.5cm}
\centerline{\large EUROPEAN ORGANIZATION FOR NUCLEAR RESEARCH (CERN)}
\vspace*{1.5cm}
\noindent
\begin{tabular*}{\linewidth}{lc@{\extracolsep{\fill}}r@{\extracolsep{0pt}}}
\ifthenelse{\boolean{pdflatex}}%
{\vspace*{-1.5cm}\mbox{\!\!\!\includegraphics[width=.14\textwidth]{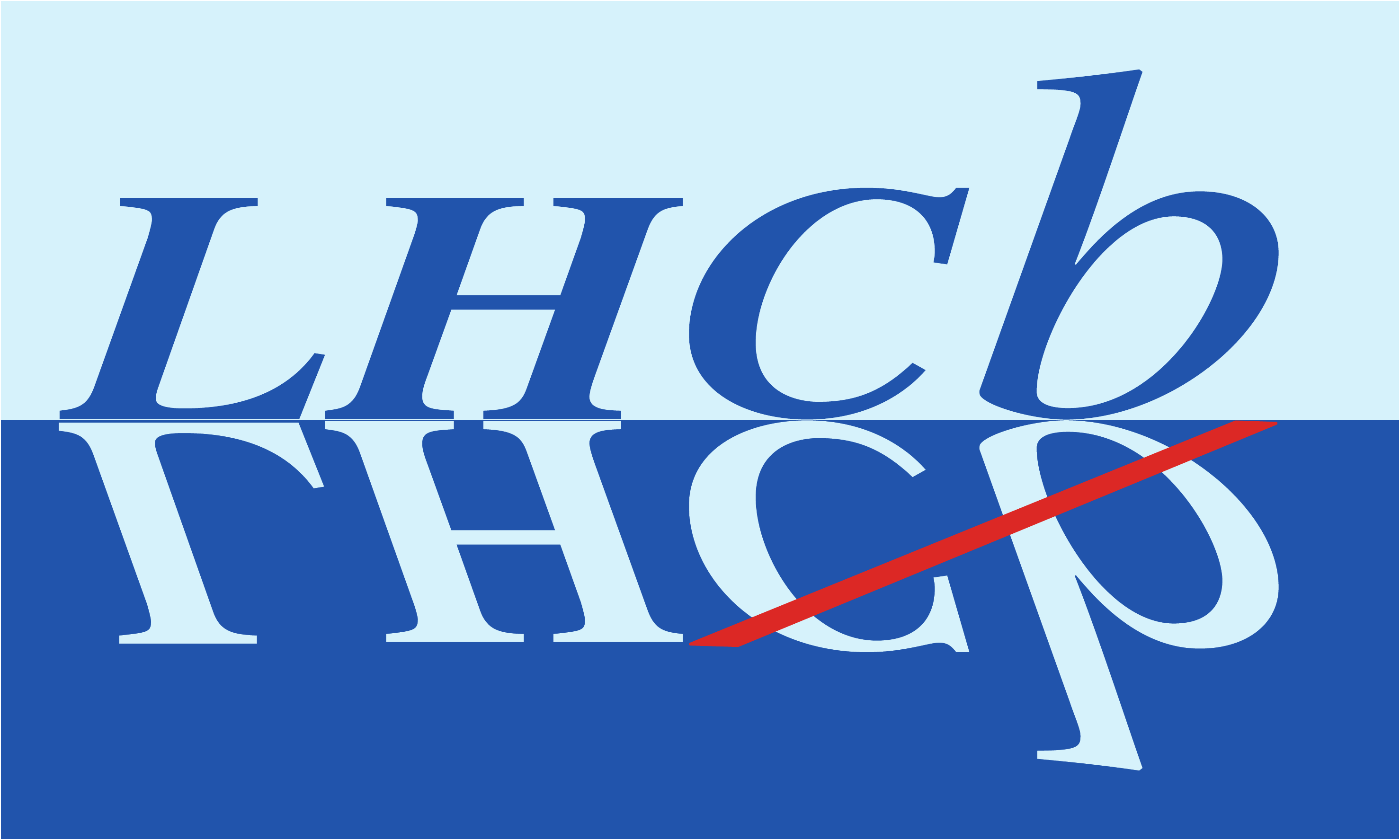}} & &}%
{\vspace*{-1.2cm}\mbox{\!\!\!\includegraphics[width=.12\textwidth]{figs/lhcb-logo.eps}} & &}%
\\
 & & CERN-EP-2024-340 \\  %
 & & LHCb-PAPER-2024-051 \\  %
 & &February 6, 2025 \\ %
 & & \\
\end{tabular*}

\vspace*{3.0cm}

{\normalfont\bfseries\boldmath\huge
\begin{center}
  \papertitle 
\end{center}
}

\vspace*{0.5cm}

\begin{center}
\paperauthors\footnote{Authors are listed at the end of this paper.}
\end{center}

\vspace{\fill}

\begin{abstract}
\noindent
The first measurement of the $C\!P$ asymmetry of the decay rate ($A_{C\!P}$) and the $C\!P$ average ($\Sigma A_{\text{FB}}$) and $C\!P$ asymmetry ($\Delta A_{\text{FB}}$) of the forward-backward asymmetry in the muon system of $\mathit{\Lambda}^+_c\to p\mu^+\mu^-$  decays is reported. The measurement is performed using a data sample of proton-proton collisions, recorded by the LHCb experiment from 2016 to 2018 at a center-of-mass energy of 13$\aunit{TeV}$, which corresponds to an integrated luminosity of 5.4$\aunit{fb}^{-1}$. The asymmetries are measured in two regions of dimuon mass near the $\phi$-meson mass peak. The dimuon-mass integrated results are
\begin{align*}
A_{C\!P} &= (-1.1 \pm 4.0 \pm 0.5)\%,\\
\Sigma A_{\text{FB}} &= (\phantom{-}3.9 \pm 4.0  \pm 0.6)\%,\\
\Delta A_{\text{FB}} &= (\phantom{-}3.1 \pm 4.0  \pm 0.4)\%,
\end{align*}
where the first uncertainty is statistical and the second systematic. The results are consistent with the conservation of $C\!P$ symmetry and the Standard Model expectations.
\end{abstract}

\vspace*{2.0cm}

\begin{center}
 Published in Phys. Rev. D111 (2025) L091102
\end{center}

\vspace{\fill}

{\footnotesize 
\centerline{\copyright~\papercopyright. \href{\paperlicenceurl}{\paperlicence}.}}
\vspace*{2mm}

\end{titlepage}

\newpage
\setcounter{page}{2}
\mbox{~}
%
%
%
%
%
%
%
%
%
%
%
%
%
%
%
%
%
%
%
%

 %
%

\renewcommand{\thefootnote}{\arabic{footnote}}
\setcounter{footnote}{0}

\cleardoublepage

\pagestyle{plain} %
\setcounter{page}{1}
\pagenumbering{arabic}

%
%
%

%
%
%
Rare decays that are sensitive to transitions between \cquark and \uquark quarks in association with the simultaneous emission of a pair of oppositely charged leptons ($\ell^+\ell^-$) offer the opportunity to explore flavor-changing neutral-currents (FCNCs) in the up-type quark sector. 
In the Standard Model (SM), FCNC transitions are only generated by loop-level processes and suppressed by the Glashow--Iliopoulos--Maiani (GIM) mechanism~\cite{GIM}. In the charm system, the GIM mechanism leads to a particularly strong suppression with respect to the down-type quark sector. Thus, studies of rare charm decays are sensitive probes of beyond-Standard-Model phenomena and constitute a complementary testing ground with respect to studies of rare beauty and strange hadron decays. New particles and interactions extending the SM can lead to modifications of branching fractions, modify angular distributions of final-state particles, or introduce additional sources of charge-parity (\CP) asymmetry~\cite{Fajfer:2005ke,Bigi:2012,Fajfer:2012nr,Cappiello,Paul:2012ab,Fajfer:2015mia,deBoer:2015boa,deBoer:2018,Bause:2019vpr,Bause:2020obd,Gisbert:2020vjx,Bharucha:2020eup,Fajfer:2023tkp,Mandal:2024lbp}. The \lhcb collaboration has previously succeeded in measuring branching fractions of rare charm meson decays at the level of $10^{-7}$ and recently published the first angular analysis and a search for \CP violation in rare decays of neutral \Dz mesons~\cite{LHCb-PAPER-2017-019,LHCb-PAPER-2018-020,LHCb-PAPER-2021-035,LHCb-PAPER-2024-047}. However, measurements of rare baryonic charm decays comprising two charged leptons in the final state remain far less explored in the experimental landscape~\cite{KODAMA199585,BaBar:2011ouc,LHCb-PAPER-2024-005}.

The decays of $\Lc$ baryons to \mbox{$\proton \mu^+ \mu^-$} final states\footnote{Charge-conjugate decays are implied throughout this letter unless stated otherwise.} proceed at short distances via \mbox{$c \to u \ell^+ \ell^-$} transitions, which in the SM lead to branching fractions below $\mathcal{O}(10^{-8})$~\cite{Meinel:2017ggx,Faustov:2018dkn,golz2022probing,Gisbert:2024kob,Zhang:2024asb}. However, the total decay width is dominated by intermediate resonant contributions of the form $\Lc \to \proton X(\to \mu^+ \mu^-)$, where $X$ can be a short-lived $\eta$, $\rhoz$, $\omegaz$ or $\phi$ meson that subsequently decays into two muons. These long-distance contributions increase the branching fraction to $\mathcal{O}(10^{-6})$. Recently, the \lhcb collaboration has published an updated measurement of the branching fractions of $\Lc \to p \mu^+ \mu^-$ decays in different regions of the dimuon mass, $m(\mu^+\mu^-)$~\cite{LHCb-PAPER-2024-005}. The measurement is performed relative to the branching fraction in the $m(\mu^+\mu^-)$ region around the known $\phi$-meson mass~\cite{PDG2024}, which is the dominant contribution to the total decay rate. An upper limit on the branching fraction \mbox{$\BF(\Lc \to \proton \mu^+ \mu^-) < 2.9 \times 10^{-8}$} at 90\% confidence level is set in regions of $m(\mu^+\mu^-)$ where the influence of intermediate resonances is minimal and sensitivity to beyond-SM contributions is largest. 
Further separation of short and long-distance contributions can be reached by studying angular distributions and \CP asymmetries, which complement searches for new phenomena in decays of \Dz and \Dp mesons because of the nonzero spin of the \Lc baryons\cite{golz2022probing,Gisbert:2024kob,Zhang:2024asb}. However, to date, no measurements of \CP or angular asymmetries in \Lcpmm decays have been performed.

This letter presents the first measurement of the direct \CP asymmetry, as well as the \CP average and \CP asymmetry of the forward-backward asymmetry in the lepton system, in \Lcpmm decays. 
The analysis uses  proton-proton (\proton\proton) collision data recorded  by the LHCb experiment at a center-of-mass energy of 13\tev in the years 2016, 2017 and 2018, which correspond to an integrated luminosity of 5.4\invfb. 
The analysis uses \Lc baryons produced directly in the primary \proton\proton interaction and the same data as used in Ref.~\cite{LHCb-PAPER-2024-005}. However, the signal candidate selection is reoptimized for the measurement of asymmetries. The \CP asymmetry, \Acp, is defined in terms of the difference of decay rates for $\Lc$ and $\Lcbar$ decays to $\proton \mup \mun$ and $\antiproton \mup \mun$ final states as
\begin{equation}
    \Acp \equiv \frac{\Gamma(\Lc \to \proton \mup \mun) - \Gamma(\Lcbar \to \antiproton \mup \mun)}{\Gamma(\Lc \to \proton \mup \mun) + \Gamma(\Lcbar \to \antiproton \mup \mun)},
\end{equation}
where $\Gamma$ is the decay rate. The forward-backward asymmetry, \AFB, is defined as 
\begin{equation}
    \AFB  \equiv \frac{\Gamma( \cos\theta>0) - \Gamma(\cos\theta<0)}{\Gamma(\cos\theta>0) + \Gamma(\cos\theta<0)},
\end{equation}
where $\theta$ is the angle between the direction of the positively charged lepton in the dimuon rest frame and the flight direction of the dimuon system in the rest frame of the $\Lc$ baryon. In contrast, for $\Lcbar$ baryons the angle is measured relative to the flight direction of the negatively charged lepton. The forward-backward asymmetry is measured separately for \Lc and \Lcbar baryons and referred to as \AFBLc and \AFBLcbar, respectively, which allows the \CP average, $\Sigma \AFBcp$, and the \CP asymmetry, $\Delta \AFBcp$, to be defined as
\begin{align}
\begin{split}
    \Sigma \AFBcp &\equiv 1/2 \cdot \left[\AFBLc + \AFBLcbar\right],\\
    \Delta \AFBcp &\equiv 1/2 \cdot \left[\AFBLc - \AFBLcbar\right]. 
\end{split}\label{eq:deltaSigmaA}
\end{align} 
These observables further enhance the sensitivity of the measurement to the real and imaginary parts of beyond-SM couplings~\cite{Gisbert:2024kob}. 

Given the current experimental sensitivity, the asymmetries measured in this letter are null tests of the SM. Interference effects between SM resonant and beyond-SM amplitudes can produce asymmetries as large as $\mathcal{O}(\%)$~\cite{golz2022probing}, referred to as \textit{resonance enhanced} asymmetries.
The measurement is performed in the $m(\mu^+\mu^-)$ region which is dominated by the intermediate $\phi$ meson, where the signal yield is sufficiently high to measure \CP and angular asymmetries. To be sensitive to variations of the asymmetries over the phase space, the observables are determined in two regions, which are defined symmetrically below and above the known mass of the $\phiz$ meson\cite{PDG2024}.

The \lhcb detector~\cite{LHCb-DP-2008-001,LHCb-DP-2014-002} is a single-arm forward spectrometer designed for the study of particles containing \bquark or \cquark quarks. It includes a high-precision tracking system consisting of a silicon-strip vertex detector surrounding the \proton\proton interaction region, a large-area silicon-strip detector located upstream of a dipole magnet with a bending power of about 4\,T\,m, and three stations of silicon-strip detectors and straw drift tubes placed downstream of the magnet. The polarity of the magnetic field is reversed periodically throughout the data-taking. Particle identification is provided by two ring-imaging Cherenkov detectors, an electromagnetic and a hadronic calorimeter, and a muon system composed of alternating layers of iron and multiwire proportional chambers. Events are selected online by a trigger that consists of a hardware stage, which is based on information from the calorimeter and muon systems, followed by a software stage which performs a full event reconstruction~\cite{LHCb-PAPER-2019-001}. Simulation is required to model the effects of the detector acceptance, the imposed selection requirements and to describe the signal and backgrounds.
In the simulation, $pp$ collisions are generated using \pythia~\cite{Sjostrand:2007gs,*Sjostrand:2006za} with a specific \lhcb configuration~\cite{LHCb-PROC-2010-056}. Decays of unstable particles are described by \evtgen~\cite{Lange:2001uf}, in which final-state radiation is generated using \photos~\cite{davidson2015photos}. The interaction of the generated particles with the detector, and its response, are implemented using the \geant toolkit~\cite{Allison:2006ve, *Agostinelli:2002hh} as described in Ref.~\cite{LHCb-PROC-2011-006}. The underlying $pp$ interaction is reused multiple times, with an independently generated signal decay for each~\cite{LHCb-DP-2018-004}.

The hardware trigger requires the presence of a muon with large transverse momentum, $\pt$, which is compatible with one of the two muons of the signal candidate. Furthermore, candidates where a positive trigger decision is caused by the presence of muons with high $\pt$, or a hadron, photon or electron with high transverse energy in the calorimeters due to other particles produced in the \proton\proton collision, are also considered.
The subsequent software trigger selects in a first stage events where a pair of tracks satisfies a multivariate classifier based on geometric and kinematic criteria that identifies candidates consistent with the displaced decay of a charm hadron, or alternatively events with a reconstructed dimuon vertex which is displaced from any primary vertex (PV). In the second software-trigger stage, candidate \Lc baryons are constructed by combining three charged tracks that form a good-quality secondary vertex. Each of the tracks is required to be inconsistent with originating from a PV, to satisfy $\pt>0.3\gevc$ and to have a minimum momentum $p$ of $3\gevc$. The scalar sum of the $\pt$ of the three tracks has to exceed $0.5\gevc$. The \Lc candidate must have a decay vertex significantly separated from any PV, and compatible with originating from one of the PVs. Therefore, only \Lc candidates with a reconstructed momentum vector that aligns with the vector connecting the secondary and primary vertices are further considered. 

Additional selection criteria are applied offline. The reconstructed mass of the \Lc candidates, $\mLc$, is limited to the range $[2147,2486]\mevcc$. Stringent particle identification criteria are placed on the particles identified as protons to suppress background from misidentified $\Dp \to \pip \mup\mun$ decays, where the pion is wrongly identified as a proton. The selection is further optimized to reduce the contributions of two major sources of background. The first comprises random associations of unrelated tracks that coincidentally fulfill the selection criteria. To reduce this combinatorial background, a multivariate selection based on a boosted decision tree (BDT) classifier~\cite{Breiman,Roe} as implemented in the TMVA toolkit~\cite{TMVA} is employed. The BDT classifier is trained using simulated candidates as proxies for the signal, and data candidates with $m(\proton\mup\mun)<2256.5\mevcc$ or $m(\proton\mup\mun)>2321.5\mevcc$ for the background.
The AdaBoost algorithm~\cite{AdaBoost} is used when training the BDT classifier in ten disjoint training subsamples, which is subsequently applied using a \textit{k-}fold cross-validation~\cite{Blum1999BeatingTH}.  The choice of features that are used in the training has been based on the selection described in Ref.\cite{LHCb-PAPER-2024-005} and comprises kinematic and topological features of the \Lc candidates, as well as features related to the isolation of the signal candidates with respect to other tracks in the event. The variables are chosen to minimize the correlation with the \Lc mass to avoid artificial sculpting of the mass spectrum of \Lc candidates.

The second major background arises from hadronic $\Lc$ baryon decays to $\proton \pip \pim$ final states, where two oppositely charged pions are identified as muons. Despite the low pion to muon misidentification probability\cite{LHCb-DP-2018-001}, the large branching fraction of $\Lc \to \proton \pip \pim$ decays~\cite{PDG2024} causes a peaking background. This is further suppressed by a multivariate muon-identification discriminant that combines the information from the Cherenkov detectors, the calorimeters and the muon chambers.

The threshold on the muon-identification discriminant is simultaneously optimized with the requirement on the BDT classifier output by minimizing the uncertainty on the asymmetries. This is achieved by randomly splitting the data samples into two halves, simulating a vanishing asymmetry in the samples as expected in null-test measurements.

If after the full selection an event contains more than one \Lc candidate, only one is randomly selected. This requirement removes less than one percent of candidates. To avoid potential biases on the measured quantities, all asymmetries were shifted by a random offset and only examined after the analysis procedure had been finalized.

The reconstruction and selection requirements result in efficiency losses that correlate with the decay kinematics and vary as a function of $m(\mu^+\mu^-)$ and $\cos \theta$. A correction for relative efficiency variations across the phase space is applied using simulated samples to ensure an unbiased determination of the measured asymmetries. The simulation samples are corrected for known differences with respect to data in particle identification response~\cite{LHCb-DP-2018-001}. A two-dimensional correction map is employed to assign weights to each candidate that account for relative efficiency variations and can be found in the supplemental material~\cite{supplemental}. 

To determine \Acp, the so-called \textit{raw} asymmetry, \Acpraw, is measured, which also receives contributions from nuisance asymmetries of approximately $4\%$. These asymmetries are caused by the different production cross-sections of \Lc and \Lcbar baryons, quantified as the production asymmetry, $\Aprod(\Lc)$, and a nonequal detection efficiency of protons and antiprotons which leads to a detection asymmetry, $\Adet(\proton)$. The detection asymmetry from the pair of oppositely charged muons is estimated to be negligible. For small asymmetries, the raw asymmetry can be approximated as
\begin{equation}
\begin{aligned} 
\Acpraw(\Lc \to \proton \mu^+\mu^-)  &\equiv \frac{N(\Lc \to \proton \mup \mun) - N(\Lcbar \to \antiproton \mup \mun)}{N(\Lc \to \proton \mup \mun) + N(\Lcbar \to \antiproton \mup \mun)} \\
            &\approx \Acp(\Lc \to \proton \mu^+\mu^-) + \Aprod(\Lc) + \Adet(\proton)
\end{aligned}
\label{eq:ACPRAW}
\end{equation}
The \CP asymmetry is obtained by subtracting the raw asymmetry measured in a control data sample of $\Lc \to \proton \KS$ decays, where effects of \CP violation are assumed to be negligible and which is subject to the same nuisance asymmetries, as 
\begin{equation}
    \Acp(\Lc \to \proton \mu^+\mu^-) = \Acpraw(\Lc \to \proton \mu^+\mu^-) - \Acpraw(\Lc \to \proton \KS).
\end{equation}
Since $\Adet(\proton)$ and $\Aprod(\Lc)$ depend on the kinematics of the \proton and \Lc particles, a weighting scheme is developed to match the final-state kinematics of the control mode to that of the signal mode using a BDT algorithm with gradient boosting~\cite{Breiman,Roe,TMVA}. 

\begin{figure}[tb]
\centering
\includegraphics[width=.6\textwidth]{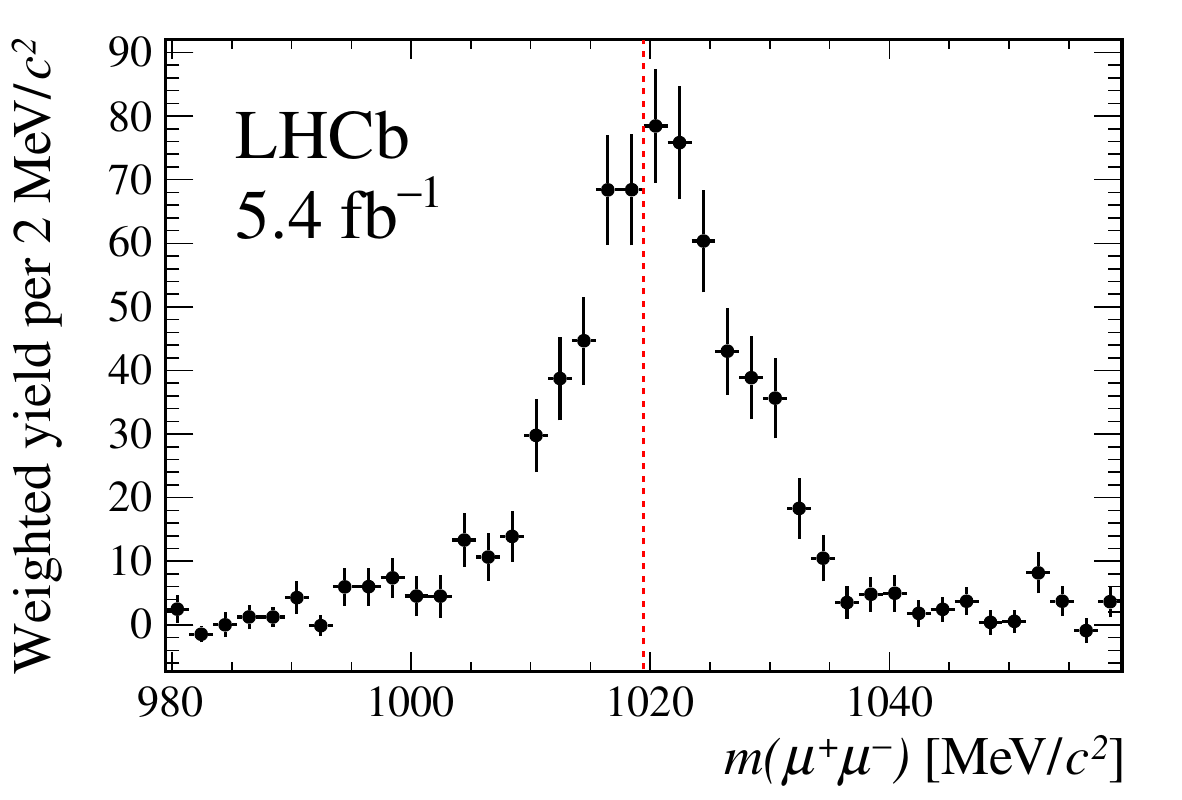}\hfil
\caption{Background-subtracted distribution of dimuon mass for $\mathit{\Lambda}^+_c\to p\mu^+\mu^-$ candidates. The red line indicates the known mass of the $\phi$ meson, which marks the boundary between the two considered dimuon-mass regions. Weights are assigned to correct for efficiency variations as described in the main text.}\label{fig:mmumu}
\end{figure}

To measure the asymmetries, the data are split in four disjoint subsamples, which are defined by the flavor of the \Lc baryon and the sign of $\cos \theta$. The measurement is performed separately in the two \mmumu regions $\phil$ and  $\phih$, defined as $[979.46,1019.46]\mevcc$ and $[1019.46,1059.46]\mevcc$. In the following, these dimuon-mass regions refer to ranges in the reconstructed dimuon mass, without correcting for effects due to a finite experimental mass resolution of approximately 8\mevcc\cite{supplemental}. Figure~\ref{fig:mmumu} shows the background-subtracted dimuon-mass spectrum of \Lcpmm candidates using a sideband subtraction, with the boundary between the two \mmumu ranges also indicated. 

The asymmetries are then determined through unbinned extended maximum-likelihood fits to the weighted \mLc distribution of selected candidates, where the weights correct for the phase-space-dependent efficiency variations. The asymmetries and yields are determined from simultaneous fits, in which they are treated as free parameters. The total fit function consists of three components: the \Lcpmm signal, combinatorial background and background from misidentified \Lcppp decays. The signal component is described by the sum of a Johnson $S_U$ function~\cite{Johnson:1949zj} and a Gaussian function, to account for asymmetric tails of the distribution. The parameters of the model are determined by a fit to simulated candidates which are selected with the same requirements as the data candidates. The peak position of the distribution is allowed to vary in the fit to the data separately for the two dimuon-mass regions and baryon flavors. The mass shape of the combinatorial background is parameterized using a fourth-order Chebyshev polynomial with parameters determined by a fit to a control data sample, where all three final-state particles carry the same electric charge. These candidates are entirely made of random combinations of tracks and used as a proxy for the combinatorial background. The misidentified \Lcppp background is also described by a Johnson $S_U$ function~\cite{Johnson:1949zj} whose parameters are determined using simulated candidates of \Lcppp decays assigning the muon mass hypothesis to both pions. Due to limited simulation sample sizes, not all selection criteria can be applied and the requirement on muon particle identification is applied to only one of the two pions. The \CP and forward-backward asymmetries of misidentified background are determined using high-yield data control samples of \Lcppp decays and are fixed in the baseline fit that is used to measure the asymmetries. Assumptions in the background modeling give rise to systematic uncertainties which are described later in this letter.

\begin{figure}[t]
\centering
\includegraphics[width=.7\textwidth]{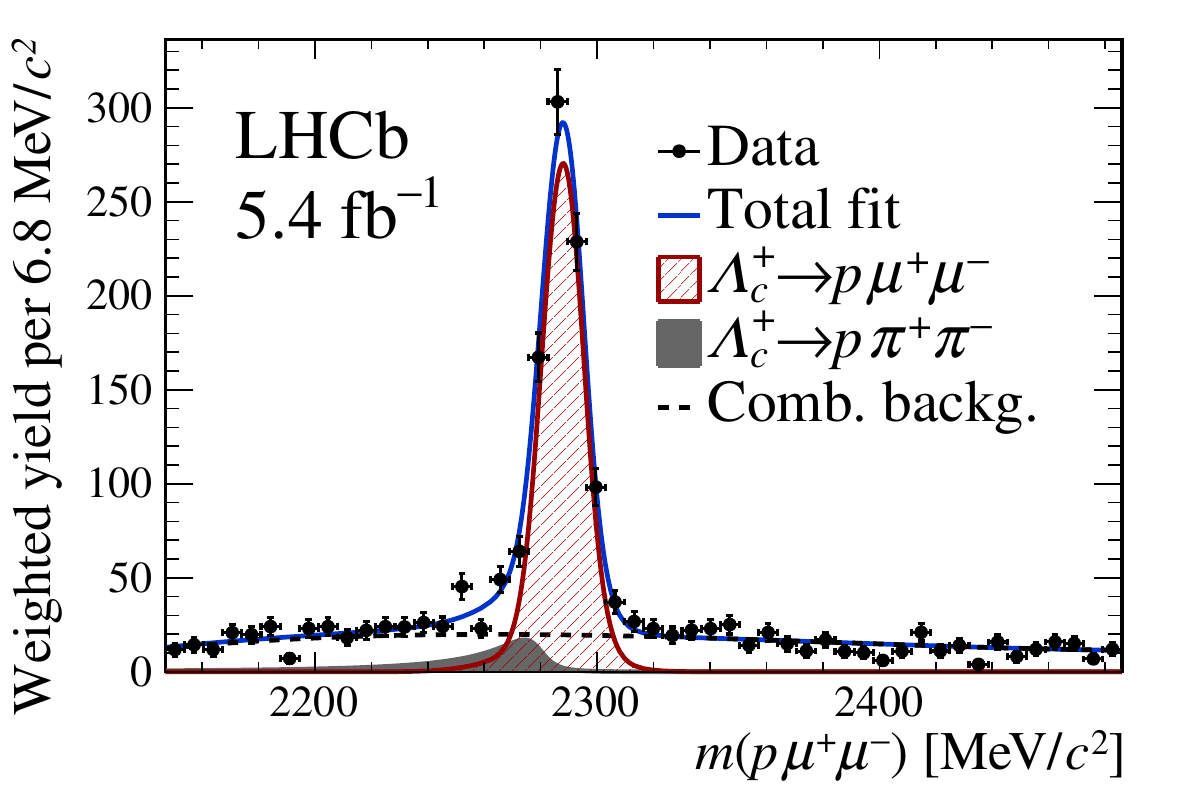}\hfil
\caption{Distribution of $m(p\mu^+\mu^-)$ for efficiency-weighted candidates, together with the fit projection.}\label{fig:FullDataFits}
\end{figure}

In the fit, the free parameters are the yields of each fit component, the asymmetries of signal and combinatorial background, and the signal peak position. The fits are performed separately in the two dimuon-mass regions. The fits are validated to return unbiased estimates of the asymmetries and their uncertainties using large samples of pseudoexperiments. The resulting signal and background yields, as well as the asymmetries \Acp, \AFBLc and \AFBLcbar are listed in Table~\ref{table:yieldsandasy}. The mass spectrum of selected \Lcpmm candidates is shown in Fig.~\ref{fig:FullDataFits}, together with the fit projection. 

\begin{table*}[t]
\centering
\caption{Yields after correcting for relative efficiency variations and measured asymmetries for \Lcpmm decays in the two dimuon-mass regions. For the asymmetries the first uncertainty is statistical and the second systematic.}\label{table:yieldsandasy}
\resizebox{
\ifdim\width>\textwidth
	\textwidth
\else
	\width
\fi}{!}{
\ifthenelse{\boolean{wordcount}}{}{%
\begin{tabular}{c r@{\,$\pm$\,}l r@{\,$\pm$\,}l r@{\,$\pm$\,}l r@{\,$\pm$\,}c@{\,$\pm$\,}l r@{\,$\pm$\,}c@{\,$\pm$\,}l r@{\,$\pm$\,}c@{\,$\pm$\,}l}
\toprule
 & \multicolumn{6}{c}{Efficiency-weighted yields} & \multicolumn{9}{c}{Asymmetries}\\
{$\mmumu$} & \multicolumn{2}{c}{Signal} & \multicolumn{2}{c}{Misid.~back.} & \multicolumn{2}{c}{Comb.~back.} & \multicolumn{3}{c}{$\Acp$ [\%]} & \multicolumn{3}{c}{$\AFBLc$ [\%]} & \multicolumn{3}{c}{$\AFBLcbar$ [\%]} \\
\midrule
\phil & 346 & 22 & 57 & 21 & 437 & 26 & $-0.8$ & 6.2 & 0.6 & 11.7 & 8.5 & 1.1 & $2.2$ & 8.7 & 1.4\\
\phih & 435 & 22 & 35 & 17 & 390 & 25 & $-1.4$ & 5.3 & 0.6 & 3.5 & 7.2 & 0.9 & $-0.3$ & 7.4 & 1.1\\
\bottomrule
\end{tabular}
}}
\end{table*}

The dominant systematic uncertainties affect all asymmetries and arise from limited knowledge of the models used in the mass fits and possible imperfections in the correction for phase-space-dependent efficiency variations. 
To evaluate the systematic uncertainty related to the fit model, pseudoexperiments are performed, where alternative parametrizations of the signal and background components are tested against the baseline fit model. For the signal component, a Crystal Ball function~\cite{Skwarnicki:1986xj} and a Gaussian function are chosen as alternatives. In addition, the requirements used to select the control sample to determine the shape of the combinatorial background are varied, and an alternative shape is obtained from a fit, where the shape parameters are left free to vary when fitting \Lcpmm signal candidates. To address the inability to apply the signal muon particle-identification requirements on the simulated \Lcppp decays, an alternative shape is determined on a sample where muon particle identification requirements are not applied. The asymmetries of the misidentified background component, which are fixed in the baseline fit, are determined in control samples with modified selection requirements and varied within the observed deviations from the baseline values. The studies are performed separately for both dimuon-mass regions. For all tested alternatives, the mean bias with respect to the baseline model is determined for the measured asymmetries, and the largest observed bias over the set of shape variations in each dimuon-mass region is taken as the corresponding systematic uncertainty.

Systematic uncertainties related to the correction for efficiency variations are evaluated by creating alternative correction maps targeting aspects of the selection which might be subject to residual differences in simulation and data. The alternative maps are obtained by applying weights to the simulated candidates to match the number of reconstructed tracks and the BDT output distribution observed in data, and by placing requirements on particle identification for the final-state particles neglecting the corrections that are applied in the baseline analysis. Using the alternative correction maps, the measurement is repeated on resampled data sets which are obtained by bootstrapping the original sample~\cite{efron:1979}. The root mean square among the variations of the results that are observed with the alternative maps is taken as the systematic uncertainty. The uncertainty related to the limited simulation sample size is also propagated and considered as systematic uncertainty on the measured asymmetries.

Additionally, the following three sources of systematic uncertainty are only relevant for the measurement of \Acp. The first source arises from the instrumental asymmetry correction due to limitations of the kinematic equalization procedure. A systematic uncertainty is assigned by evaluating the effect of residual discrepancies of the kinematic distributions of signal and control mode candidates after the weighting. The root mean square among the variations of the results obtained after a second one-dimensional weighting of the \Lc and proton kinematic distributions is taken as a systematic uncertainty. The second source arises from the combined effects of \CP violation and mixing in the neutral kaon system and the different interaction probabilities of \Kz and \Kzb particles with the detector material~\cite{PhysRevLett.42.13,PhysRevLett.75.402} and a systematic uncertainty is evaluated following the method described in Refs.~\cite{LHCb-PAPER-2014-013,LHCb-PAPER-2022-024}.
Finally, the fraction of \Lc baryons arising from decays of \bquark-flavored hadrons is determined for signal and control mode candidates by studying the distribution of the impact parameter, which is defined as the minimum distance of the \Lc trajectory to the PV. The difference of these fractions for signal and control mode candidates, together with the difference in asymmetries for \Lc baryons produced in the primary interaction and from \bquark-hadron decays, are translated into a systematic uncertainty related to an imperfect cancellation of production asymmetries. 

The finite angular resolution of the \lhcb detector is relevant for the measurement of the forward-backward asymmetry, and a systematic uncertainty is assigned by estimating the fraction of candidates where the sign of $\cos \theta$ changes due to the finite detector resolution in simulations.

A summary of the systematic uncertainties, together with the correlations of the measured asymmetries due to the systematic uncertainties, can be found in the supplemental material of this letter~\cite{supplemental}. The size of the systematic uncertainty depends on the considered asymmetry and dimuon-mass region but typically constitutes between 10\% and 15\% of the corresponding statistical uncertainty.

The analysis is repeated on statistically independent data subsets to check for biases from specific instrumental effects. The criteria to split the subsets include the number of PVs and reconstructed tracks in the event, the data-taking year, the magnetic-field orientation, the trigger classification, the \Lc baryon transverse momentum and the impact parameter significance of the \Lc candidate with respect to the PV, which is defined as the difference in the vertex-fit $\chi^2$ of a given PV, reconstructed with and without considering the candidate. The resulting variations of the measured asymmetries are consistent with statistical fluctuations.

\begin{table*}[t]
\centering
\caption{\CP average and asymmetry of the forward-backward asymmetry for \Lcpmm decays in the dimuon-mass regions. The first uncertainty is statistical and the second systematic.}\label{table:results}
\resizebox{
\ifdim\width>\textwidth
	\textwidth
\else
	\width
\fi}{!}{
\ifthenelse{\boolean{wordcount}}{}{%
\begin{tabular}{c r@{\,$\pm$\,}c@{\,$\pm$\,}l r@{\,$\pm$\,}c@{\,$\pm$\,}l r@{\,$\pm$\,}c@{\,$\pm$\,}l}
\toprule
 {$\mmumu$} & \multicolumn{3}{c}{$\SAFB$ [\%]} & \multicolumn{3}{c}{$\DAFB$ [\%]} \\
\midrule
\phil & 6.9 & 6.1 & 1.0 & 4.8 & 6.1 & 0.8\\
\phih & 1.6 & 5.2 & 0.8 & 1.9 & 5.2 & 0.6\\
\bottomrule
\end{tabular}
}}
\end{table*}

\begin{figure}[tb]
\centering
\includegraphics[width=.7\textwidth]{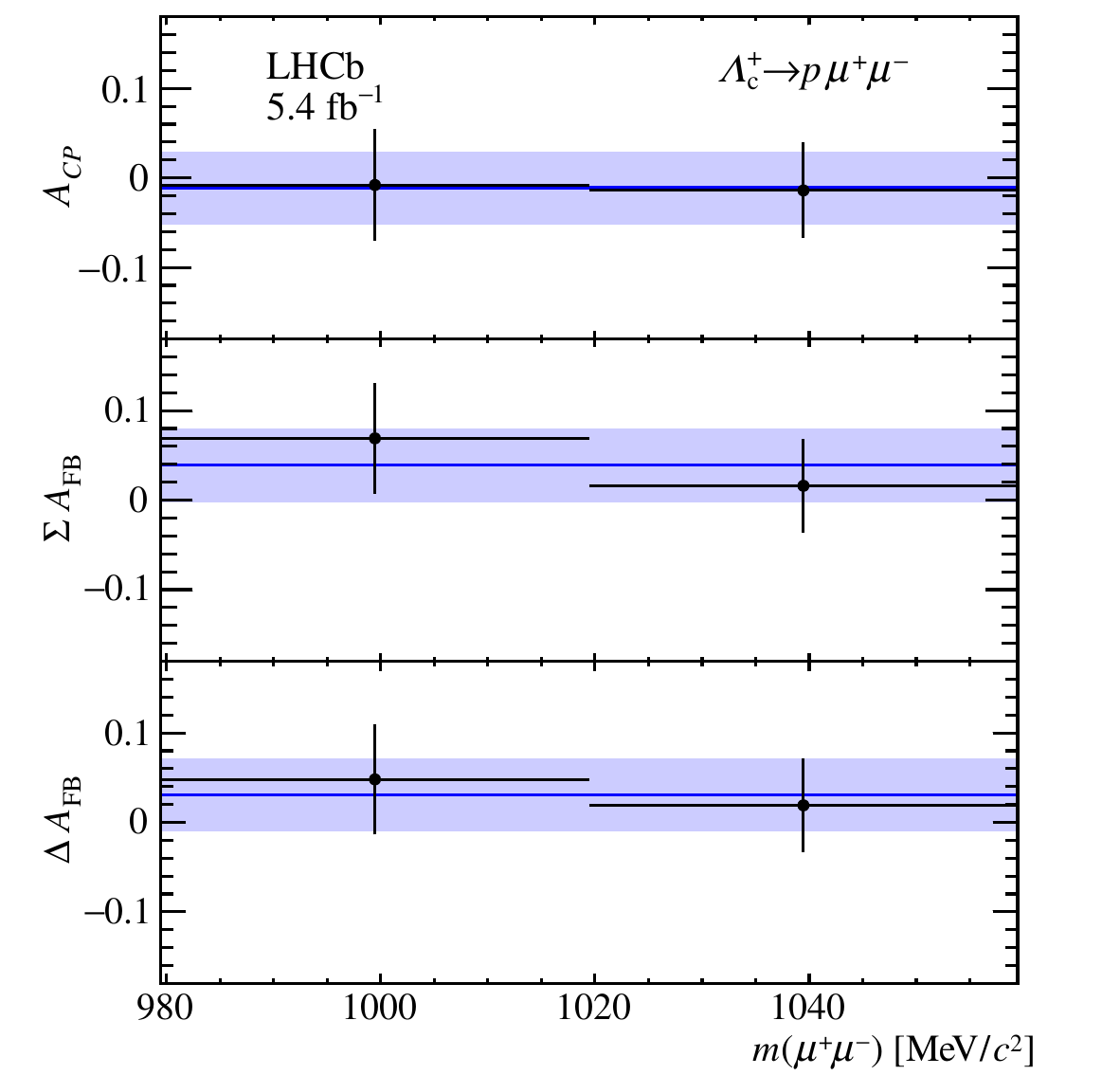}\hfil
\caption{Measured signal asymmetries in the regions of dimuon mass (black points) and their dimuon-mass-integrated value with its uncertainty (blue). The shown uncertainty comprises both the statistical and systematic uncertainty.}\label{fig:results}
\end{figure}

Using Eq.~\ref{eq:deltaSigmaA} and considering correlations between the systematic uncertainties~\cite{supplemental}, \DAFB and \SAFB are computed from the measured asymmetries listed in Table~\ref{table:yieldsandasy} and the results can be found in Table~\ref{table:results}. Correlations between the statistical uncertainties of the measured asymmetries are negligible. Furthermore, the dimuon-mass integrated values are determined, following the procedure described in Ref.~\cite{LYONS1988110}, and found to be: 
\ifthenelse{\boolean{wordcount}}{}{\integratedresults{*}}
where the first uncertainty is statistical and the second systematic.
All results are also shown in Fig.~\ref{fig:results} and are consistent with zero, confirming the SM expectation. 

In summary, a search for resonance-enhanced angular and \CP asymmetries in \Lcpmm decays in two \mmumu regions near the $\phiz$-meson mass is presented. The results are based on the analysis of \proton\proton collision data recorded by the \lhcb experiment in the years 2016, 2017 and 2018 at a center-of-mass energy of 13\tev, which correspond to an integrated luminosity of 5.4$\invfb$. The asymmetries are measured in two regions of dimuon mass to enhance the sensitivity to beyond-SM effects. The results confirm the SM prediction and will help to constrain the parameter space of models extending the SM~\cite{golz2022probing,Gisbert:2024kob}. Future data sets recorded by the upgraded \lhcb detector~\cite{LHCb-DP-2022-002} with significantly increased yields will allow the statistical uncertainties to be further reduced, and the measurement to be extended to additional dimuon-mass regions.

 
%
%
%
%
\section*{Acknowledgements}
\noindent We express our gratitude to our colleagues in the CERN
accelerator departments for the excellent performance of the LHC. We
thank the technical and administrative staff at the LHCb
institutes.
We acknowledge support from CERN and from the national agencies:
CAPES, CNPq, FAPERJ and FINEP (Brazil); 
MOST and NSFC (China); 
CNRS/IN2P3 (France); 
BMBF, DFG and MPG (Germany); 
INFN (Italy); 
NWO (Netherlands); 
MNiSW and NCN (Poland); 
MCID/IFA (Romania); 
MICIU and AEI (Spain);
SNSF and SER (Switzerland); 
NASU (Ukraine); 
STFC (United Kingdom); 
DOE NP and NSF (USA).
We acknowledge the computing resources that are provided by CERN, IN2P3
(France), KIT and DESY (Germany), INFN (Italy), SURF (Netherlands),
PIC (Spain), GridPP (United Kingdom), 
CSCS (Switzerland), IFIN-HH (Romania), CBPF (Brazil),
and Polish WLCG (Poland).
We are indebted to the communities behind the multiple open-source
software packages on which we depend.
Individual groups or members have received support from
ARC and ARDC (Australia);
Key Research Program of Frontier Sciences of CAS, CAS PIFI, CAS CCEPP, 
Fundamental Research Funds for the Central Universities, 
and Sci. \& Tech. Program of Guangzhou (China);
Minciencias (Colombia);
EPLANET, Marie Sk\l{}odowska-Curie Actions, ERC and NextGenerationEU (European Union);
A*MIDEX, ANR, IPhU and Labex P2IO, and R\'{e}gion Auvergne-Rh\^{o}ne-Alpes (France);
AvH Foundation (Germany);
ICSC (Italy); 
Severo Ochoa and Mar\'ia de Maeztu Units of Excellence, GVA, XuntaGal, GENCAT, InTalent-Inditex and Prog. ~Atracci\'on Talento CM (Spain);
SRC (Sweden);
the Leverhulme Trust, the Royal Society
 and UKRI (United Kingdom).

 
%
\section*{Data availability statement}

Data associated to the plots in this publication as well as in supplementary materials are made available on the CERN document server in~Ref.\cite{data}. 

\addcontentsline{toc}{section}{References}
\bibliographystyle{LHCb}

\begin{mcitethebibliography}{10}
\mciteSetBstSublistMode{n}
\mciteSetBstMaxWidthForm{subitem}{\alph{mcitesubitemcount})}
\mciteSetBstSublistLabelBeginEnd{\mcitemaxwidthsubitemform\space}
{\relax}{\relax}

\bibitem{GIM}
S.~L. Glashow, J.~Iliopoulos, and L.~Maiani,
  \ifthenelse{\boolean{articletitles}}{\emph{Weak interactions with
  lepton-hadron symmetry},
  }{}\href{https://doi.org/10.1103/PhysRevD.2.1285}{Phys.\ Rev.\  \textbf{D2}
  (1970) 1285}\relax
\mciteBstWouldAddEndPuncttrue
\mciteSetBstMidEndSepPunct{\mcitedefaultmidpunct}
{\mcitedefaultendpunct}{\mcitedefaultseppunct}\relax
\EndOfBibitem
\bibitem{Fajfer:2005ke}
S.~Fajfer and S.~Prelov\v{s}ek,
  \ifthenelse{\boolean{articletitles}}{\emph{{Effects of littlest Higgs model
  in rare D meson decays}},
  }{}\href{https://doi.org/10.1103/PhysRevD.73.054026}{Phys.\ Rev.\
  \textbf{D73} (2006) 054026},
  \href{http://arxiv.org/abs/hep-ph/0511048}{{\normalfont\ttfamily
  arXiv:hep-ph/0511048}}\relax
\mciteBstWouldAddEndPuncttrue
\mciteSetBstMidEndSepPunct{\mcitedefaultmidpunct}
{\mcitedefaultendpunct}{\mcitedefaultseppunct}\relax
\EndOfBibitem
\bibitem{Bigi:2012}
I.~I. Bigi and A.~Paul, \ifthenelse{\boolean{articletitles}}{\emph{{{On CP
  asymmetries in two-, three- and four-body $D$ decays}}},
  }{}\href{https://doi.org/10.1007/JHEP03(2012)021}{JHEP \textbf{03} (2012)
  021}, \href{http://arxiv.org/abs/1110.2862}{{\normalfont\ttfamily
  arXiv:1110.2862}}\relax
\mciteBstWouldAddEndPuncttrue
\mciteSetBstMidEndSepPunct{\mcitedefaultmidpunct}
{\mcitedefaultendpunct}{\mcitedefaultseppunct}\relax
\EndOfBibitem
\bibitem{Fajfer:2012nr}
S.~Fajfer and N.~Ko\v{s}nik,
  \ifthenelse{\boolean{articletitles}}{\emph{{Resonance catalyzed CP
  asymmetries in $D\to P\ell^+\ell^-$}},
  }{}\href{https://doi.org/10.1103/PhysRevD.87.054026}{Phys.\ Rev.\
  \textbf{D87} (2013) 054026},
  \href{http://arxiv.org/abs/1208.0759}{{\normalfont\ttfamily
  arXiv:1208.0759}}\relax
\mciteBstWouldAddEndPuncttrue
\mciteSetBstMidEndSepPunct{\mcitedefaultmidpunct}
{\mcitedefaultendpunct}{\mcitedefaultseppunct}\relax
\EndOfBibitem
\bibitem{Cappiello}
L.~Cappiello, O.~Cat\`{a}, and G.~D'Ambrosio,
  \ifthenelse{\boolean{articletitles}}{\emph{{Standard model prediction and new
  physics tests for $D^{0} \rightarrow h^{+}_{1}h_{2}^{-}\ell^+\ell^-$
  ($h=\pi,K$; $\ell=e,\mu)$}},
  }{}\href{https://doi.org/10.1007/JHEP04(2013)135}{JHEP \textbf{04} (2013)
  135}, \href{http://arxiv.org/abs/1209.4235}{{\normalfont\ttfamily
  arXiv:1209.4235}}\relax
\mciteBstWouldAddEndPuncttrue
\mciteSetBstMidEndSepPunct{\mcitedefaultmidpunct}
{\mcitedefaultendpunct}{\mcitedefaultseppunct}\relax
\EndOfBibitem
\bibitem{Paul:2012ab}
A.~Paul, A.~de~la Puente, and I.~I. Bigi,
  \ifthenelse{\boolean{articletitles}}{\emph{{Manifestations of warped extra
  dimension in rare charm decays and asymmetries}},
  }{}\href{https://doi.org/10.1103/PhysRevD.90.014035}{Phys.\ Rev.\
  \textbf{D90} (2014) 014035},
  \href{http://arxiv.org/abs/1212.4849}{{\normalfont\ttfamily
  arXiv:1212.4849}}\relax
\mciteBstWouldAddEndPuncttrue
\mciteSetBstMidEndSepPunct{\mcitedefaultmidpunct}
{\mcitedefaultendpunct}{\mcitedefaultseppunct}\relax
\EndOfBibitem
\bibitem{Fajfer:2015mia}
S.~Fajfer and N.~Ko\v{s}nik,
  \ifthenelse{\boolean{articletitles}}{\emph{{Prospects of discovering new
  physics in rare charm decays}},
  }{}\href{https://doi.org/10.1140/epjc/s10052-015-3801-2}{Eur.\ Phys.\ J.\
  \textbf{C75} (2015) 567},
  \href{http://arxiv.org/abs/1510.00965}{{\normalfont\ttfamily
  arXiv:1510.00965}}\relax
\mciteBstWouldAddEndPuncttrue
\mciteSetBstMidEndSepPunct{\mcitedefaultmidpunct}
{\mcitedefaultendpunct}{\mcitedefaultseppunct}\relax
\EndOfBibitem
\bibitem{deBoer:2015boa}
S.~de~Boer and G.~Hiller, \ifthenelse{\boolean{articletitles}}{\emph{{Flavour
  and new physics opportunities with rare charm decays into leptons}},
  }{}\href{https://doi.org/10.1103/PhysRevD.93.074001}{Phys.\ Rev.\
  \textbf{D93} (2016) 074001},
  \href{http://arxiv.org/abs/1510.00311}{{\normalfont\ttfamily
  arXiv:1510.00311}}\relax
\mciteBstWouldAddEndPuncttrue
\mciteSetBstMidEndSepPunct{\mcitedefaultmidpunct}
{\mcitedefaultendpunct}{\mcitedefaultseppunct}\relax
\EndOfBibitem
\bibitem{deBoer:2018}
S.~De~Boer and G.~Hiller, \ifthenelse{\boolean{articletitles}}{\emph{{Null
  tests from angular distributions in $D \to P_1 P_2 l^+l^-$, $l=e,\mu$ decays
  on and off peak}},
  }{}\href{https://doi.org/10.1103/PhysRevD.98.035041}{Phys.\ Rev.\
  \textbf{D98} (2018) 035041},
  \href{http://arxiv.org/abs/1805.08516}{{\normalfont\ttfamily
  arXiv:1805.08516}}\relax
\mciteBstWouldAddEndPuncttrue
\mciteSetBstMidEndSepPunct{\mcitedefaultmidpunct}
{\mcitedefaultendpunct}{\mcitedefaultseppunct}\relax
\EndOfBibitem
\bibitem{Bause:2019vpr}
R.~Bause, M.~Golz, G.~Hiller, and A.~Tayduganov,
  \ifthenelse{\boolean{articletitles}}{\emph{{The new physics reach of null
  tests with $D \rightarrow \pi \ell \ell $ and $D_s \rightarrow K \ell \ell $
  decays}}, }{}\href{https://doi.org/10.1140/epjc/s10052-020-7621-7}{Eur.\
  Phys.\ J.\  \textbf{C80} (2020) 65}, Erratum
  \href{https://doi.org/10.1140/epjc/s10052-020-7621-7}{ibid.\   \textbf{C81}
  (2021) 219}, \href{http://arxiv.org/abs/1909.11108}{{\normalfont\ttfamily
  arXiv:1909.11108}}\relax
\mciteBstWouldAddEndPuncttrue
\mciteSetBstMidEndSepPunct{\mcitedefaultmidpunct}
{\mcitedefaultendpunct}{\mcitedefaultseppunct}\relax
\EndOfBibitem
\bibitem{Bause:2020obd}
R.~Bause, H.~Gisbert, M.~Golz, and G.~Hiller,
  \ifthenelse{\boolean{articletitles}}{\emph{{Exploiting $CP$-asymmetries in
  rare charm decays}},
  }{}\href{https://doi.org/10.1103/PhysRevD.101.115006}{Phys.\ Rev.\
  \textbf{D101} (2020) 115006},
  \href{http://arxiv.org/abs/2004.01206}{{\normalfont\ttfamily
  arXiv:2004.01206}}\relax
\mciteBstWouldAddEndPuncttrue
\mciteSetBstMidEndSepPunct{\mcitedefaultmidpunct}
{\mcitedefaultendpunct}{\mcitedefaultseppunct}\relax
\EndOfBibitem
\bibitem{Gisbert:2020vjx}
H.~Gisbert, M.~Golz, and D.~S. Mitzel,
  \ifthenelse{\boolean{articletitles}}{\emph{{Theoretical and experimental
  status of rare charm decays}},
  }{}\href{https://doi.org/10.1142/S0217732321300020}{Mod.\ Phys.\ Lett.\
  \textbf{A36} (2021) 2130002},
  \href{http://arxiv.org/abs/2011.09478}{{\normalfont\ttfamily
  arXiv:2011.09478}}\relax
\mciteBstWouldAddEndPuncttrue
\mciteSetBstMidEndSepPunct{\mcitedefaultmidpunct}
{\mcitedefaultendpunct}{\mcitedefaultseppunct}\relax
\EndOfBibitem
\bibitem{Bharucha:2020eup}
A.~Bharucha, D.~Boito, and C.~M\'eaux,
  \ifthenelse{\boolean{articletitles}}{\emph{{Disentangling QCD and new physics
  in $D^+\to\pi^+\ell^+\ell^-$}},
  }{}\href{https://doi.org/10.1007/JHEP04(2021)158}{JHEP \textbf{04} (2021)
  158}, \href{http://arxiv.org/abs/2011.12856}{{\normalfont\ttfamily
  arXiv:2011.12856}}\relax
\mciteBstWouldAddEndPuncttrue
\mciteSetBstMidEndSepPunct{\mcitedefaultmidpunct}
{\mcitedefaultendpunct}{\mcitedefaultseppunct}\relax
\EndOfBibitem
\bibitem{Fajfer:2023tkp}
S.~Fajfer, E.~Solomonidi, and L.~Vale~Silva,
  \ifthenelse{\boolean{articletitles}}{\emph{{$S$-wave contribution to rare
  ${D}^{0}\ensuremath{\rightarrow}{\ensuremath{\pi}}^{+}{\ensuremath{\pi}}^{\ensuremath{-}}{\ensuremath{\ell}}^{+}{\ensuremath{\ell}}^{\ensuremath{-}}$
  decays in the standard model and sensitivity to new physics}},
  }{}\href{https://doi.org/10.1103/PhysRevD.109.036027}{Phys.\ Rev.\
  \textbf{D109} (2024) 3},
  \href{http://arxiv.org/abs/2312.07501}{{\normalfont\ttfamily
  arXiv:2312.07501}}\relax
\mciteBstWouldAddEndPuncttrue
\mciteSetBstMidEndSepPunct{\mcitedefaultmidpunct}
{\mcitedefaultendpunct}{\mcitedefaultseppunct}\relax
\EndOfBibitem
\bibitem{Mandal:2024lbp}
M.~Mandal, S.~Biswas, S.~Mahata, and S.~Sahoo,
  \ifthenelse{\boolean{articletitles}}{\emph{{Searching for new physics in c
  $\to \mu$ transitions with nonuniversal Z' model}},
  }{}\href{https://doi.org/10.1142/S0217751X24500635}{Int.\ J.\ Mod.\ Phys.\
  \textbf{A39} (2024) 2450063}\relax
\mciteBstWouldAddEndPuncttrue
\mciteSetBstMidEndSepPunct{\mcitedefaultmidpunct}
{\mcitedefaultendpunct}{\mcitedefaultseppunct}\relax
\EndOfBibitem
\bibitem{LHCb-PAPER-2017-019}
LHCb collaboration, R.~Aaij {\em et~al.},
  \ifthenelse{\boolean{articletitles}}{\emph{{Observation of \Dz meson decays
  to $\pip\pim\mumu$ and $\Kp\Km\mumu$ final states}},
  }{}\href{https://doi.org/10.1103/PhysRevLett.119.181805}{Phys.\ Rev.\ Lett.\
  \textbf{119} (2017) 181805},
  \href{http://arxiv.org/abs/1707.08377}{{\normalfont\ttfamily
  arXiv:1707.08377}}\relax
\mciteBstWouldAddEndPuncttrue
\mciteSetBstMidEndSepPunct{\mcitedefaultmidpunct}
{\mcitedefaultendpunct}{\mcitedefaultseppunct}\relax
\EndOfBibitem
\bibitem{LHCb-PAPER-2018-020}
LHCb collaboration, R.~Aaij {\em et~al.},
  \ifthenelse{\boolean{articletitles}}{\emph{{Measurement of angular and \CP
  asymmetries in \mbox{\decay{\Dz}{\pip\pim\mumu}} and
  \mbox{\decay{\Dz}{\Kp\Km\mumu}} decays}},
  }{}\href{https://doi.org/10.1103/PhysRevLett.121.091801}{Phys.\ Rev.\ Lett.\
  \textbf{121} (2018) 091801},
  \href{http://arxiv.org/abs/1806.10793}{{\normalfont\ttfamily
  arXiv:1806.10793}}\relax
\mciteBstWouldAddEndPuncttrue
\mciteSetBstMidEndSepPunct{\mcitedefaultmidpunct}
{\mcitedefaultendpunct}{\mcitedefaultseppunct}\relax
\EndOfBibitem
\bibitem{LHCb-PAPER-2021-035}
LHCb collaboration, R.~Aaij {\em et~al.},
  \ifthenelse{\boolean{articletitles}}{\emph{{Angular analysis of $D^0 \to
  \pi^+ \pi ^- \mu^+ \mu^-$ and $D^0 \to K^+ K ^- \mu^+ \mu^-$ decays and
  search for \CP violation}},
  }{}\href{https://doi.org/10.1103/PhysRevLett.128.221801}{Phys.\ Rev.\ Lett.\
  \textbf{128} (2022) 221801},
  \href{http://arxiv.org/abs/2111.03327}{{\normalfont\ttfamily
  arXiv:2111.03327}}\relax
\mciteBstWouldAddEndPuncttrue
\mciteSetBstMidEndSepPunct{\mcitedefaultmidpunct}
{\mcitedefaultendpunct}{\mcitedefaultseppunct}\relax
\EndOfBibitem
\bibitem{LHCb-PAPER-2024-047}
LHCb collaboration, R.~Aaij {\em et~al.},
  \ifthenelse{\boolean{articletitles}}{\emph{{Search for $D^0$ meson decays to
  $\pi^+ \pi^- e^+ e^-$ and $K^+ K^- e^+ e^-$ final states}},
  }{}\href{https://doi.org/10.1103/PhysRevD.111.L091101}{{Phys.\ Rev.\ }
  \textbf{{D111}} (2025) L091101},
  \href{http://arxiv.org/abs/2412.09414}{{\normalfont\ttfamily
  arXiv:2412.09414}}\relax
\mciteBstWouldAddEndPuncttrue
\mciteSetBstMidEndSepPunct{\mcitedefaultmidpunct}
{\mcitedefaultendpunct}{\mcitedefaultseppunct}\relax
\EndOfBibitem
\bibitem{KODAMA199585}
K.~Kodama {\em et~al.}, \ifthenelse{\boolean{articletitles}}{\emph{Upper limits
  for charm hadron decays to two muons plus hadrons},
  }{}\href{https://doi.org/https://doi.org/10.1016/0370-2693(94)01610-O}{Phys.\
  Lett.\  \textbf{B345} (1995) 85}\relax
\mciteBstWouldAddEndPuncttrue
\mciteSetBstMidEndSepPunct{\mcitedefaultmidpunct}
{\mcitedefaultendpunct}{\mcitedefaultseppunct}\relax
\EndOfBibitem
\bibitem{BaBar:2011ouc}
BaBar collaboration, J.~P. Lees {\em et~al.},
  \ifthenelse{\boolean{articletitles}}{\emph{{Searches for rare or forbidden
  semileptonic charm decays}},
  }{}\href{https://doi.org/10.1103/PhysRevD.84.072006}{Phys.\ Rev.\
  \textbf{D84} (2011) 072006},
  \href{http://arxiv.org/abs/1107.4465}{{\normalfont\ttfamily
  arXiv:1107.4465}}\relax
\mciteBstWouldAddEndPuncttrue
\mciteSetBstMidEndSepPunct{\mcitedefaultmidpunct}
{\mcitedefaultendpunct}{\mcitedefaultseppunct}\relax
\EndOfBibitem
\bibitem{LHCb-PAPER-2024-005}
LHCb collaboration, R.~Aaij {\em et~al.},
  \ifthenelse{\boolean{articletitles}}{\emph{{Search for the rare decay of
  charmed baryon $\Lambda_c^+$ into the $p \mu^+ \mu^-$ final state}},
  }{}\href{https://doi.org/10.1103/PhysRevD.110.052007}{{Phys.\ Rev.\ }
  \textbf{{D110}} (2024) 052007},
  \href{http://arxiv.org/abs/2407.11474}{{\normalfont\ttfamily
  arXiv:2407.11474}}\relax
\mciteBstWouldAddEndPuncttrue
\mciteSetBstMidEndSepPunct{\mcitedefaultmidpunct}
{\mcitedefaultendpunct}{\mcitedefaultseppunct}\relax
\EndOfBibitem
\bibitem{Meinel:2017ggx}
S.~Meinel, \ifthenelse{\boolean{articletitles}}{\emph{{$\Lambda_c \to N$ form
  factors from lattice QCD and phenomenology of $\Lambda_c \to n \ell^+
  \nu_\ell$ and $\Lambda_c \to p \mu^+ \mu^-$ decays}},
  }{}\href{https://doi.org/10.1103/PhysRevD.97.034511}{Phys.\ Rev.\
  \textbf{D97} (2018) 034511},
  \href{http://arxiv.org/abs/1712.05783}{{\normalfont\ttfamily
  arXiv:1712.05783}}\relax
\mciteBstWouldAddEndPuncttrue
\mciteSetBstMidEndSepPunct{\mcitedefaultmidpunct}
{\mcitedefaultendpunct}{\mcitedefaultseppunct}\relax
\EndOfBibitem
\bibitem{Faustov:2018dkn}
R.~N. Faustov and V.~O. Galkin,
  \ifthenelse{\boolean{articletitles}}{\emph{{Rare $\Lambda _c\rightarrow p
  \ell ^+\ell ^-$ decay in the relativistic quark model}},
  }{}\href{https://doi.org/10.1140/epjc/s10052-018-6010-y}{Eur.\ Phys.\ J.\
  \textbf{C78} (2018) 527},
  \href{http://arxiv.org/abs/1805.02516}{{\normalfont\ttfamily
  arXiv:1805.02516}}\relax
\mciteBstWouldAddEndPuncttrue
\mciteSetBstMidEndSepPunct{\mcitedefaultmidpunct}
{\mcitedefaultendpunct}{\mcitedefaultseppunct}\relax
\EndOfBibitem
\bibitem{golz2022probing}
M.~Golz, G.~Hiller, and T.~Magorsch,
  \ifthenelse{\boolean{articletitles}}{\emph{{Probing for new physics with rare
  charm baryon (\ensuremath{\Lambda}$_{c}$, \ensuremath{\Xi}$_{c}$,
  \ensuremath{\Omega}$_{c}$) decays}},
  }{}\href{https://doi.org/10.1007/JHEP09(2021)208}{JHEP \textbf{09} (2021)
  208}, \href{http://arxiv.org/abs/2107.13010}{{\normalfont\ttfamily
  arXiv:2107.13010}}\relax
\mciteBstWouldAddEndPuncttrue
\mciteSetBstMidEndSepPunct{\mcitedefaultmidpunct}
{\mcitedefaultendpunct}{\mcitedefaultseppunct}\relax
\EndOfBibitem
\bibitem{Gisbert:2024kob}
H.~Gisbert, G.~Hiller, and D.~Suelmann,
  \ifthenelse{\boolean{articletitles}}{\emph{{Effective field theory analysis
  of rare $|\Delta c|=|\Delta u|=1$ charm decays}},
  }{}\href{https://doi.org/10.1007/JHEP12(2024)102}{JHEP \textbf{12} (2024)
  102}, \href{http://arxiv.org/abs/2410.00115}{{\normalfont\ttfamily
  arXiv:2410.00115}}\relax
\mciteBstWouldAddEndPuncttrue
\mciteSetBstMidEndSepPunct{\mcitedefaultmidpunct}
{\mcitedefaultendpunct}{\mcitedefaultseppunct}\relax
\EndOfBibitem
\bibitem{Zhang:2024asb}
S.-Q. Zhang and C.-F. Qiao, \ifthenelse{\boolean{articletitles}}{\emph{{Rare
  \ensuremath{\Lambda}c decays and new physics effects}},
  }{}\href{https://doi.org/10.1103/PhysRevD.110.114040}{Phys.\ Rev.\
  \textbf{D110} (2024) 114040},
  \href{http://arxiv.org/abs/2411.15857}{{\normalfont\ttfamily
  arXiv:2411.15857}}\relax
\mciteBstWouldAddEndPuncttrue
\mciteSetBstMidEndSepPunct{\mcitedefaultmidpunct}
{\mcitedefaultendpunct}{\mcitedefaultseppunct}\relax
\EndOfBibitem
\bibitem{PDG2024}
Particle Data Group, S.~Navas {\em et~al.},
  \ifthenelse{\boolean{articletitles}}{\emph{{\href{http://pdg.lbl.gov/}{Review
  of particle physics}}},
  }{}\href{https://doi.org/10.1103/PhysRevD.110.030001}{Phys.\ Rev.\
  \textbf{D110} (2024) 030001}\relax
\mciteBstWouldAddEndPuncttrue
\mciteSetBstMidEndSepPunct{\mcitedefaultmidpunct}
{\mcitedefaultendpunct}{\mcitedefaultseppunct}\relax
\EndOfBibitem
\bibitem{LHCb-DP-2008-001}
LHCb collaboration, A.~A. Alves~Jr.\ {\em et~al.},
  \ifthenelse{\boolean{articletitles}}{\emph{{The \lhcb detector at the LHC}},
  }{}\href{https://doi.org/10.1088/1748-0221/3/08/S08005}{JINST \textbf{3}
  (2008) S08005}\relax
\mciteBstWouldAddEndPuncttrue
\mciteSetBstMidEndSepPunct{\mcitedefaultmidpunct}
{\mcitedefaultendpunct}{\mcitedefaultseppunct}\relax
\EndOfBibitem
\bibitem{LHCb-DP-2014-002}
LHCb collaboration, R.~Aaij {\em et~al.},
  \ifthenelse{\boolean{articletitles}}{\emph{{LHCb detector performance}},
  }{}\href{https://doi.org/10.1142/S0217751X15300227}{Int.\ J.\ Mod.\ Phys.\
  \textbf{A30} (2015) 1530022},
  \href{http://arxiv.org/abs/1412.6352}{{\normalfont\ttfamily
  arXiv:1412.6352}}\relax
\mciteBstWouldAddEndPuncttrue
\mciteSetBstMidEndSepPunct{\mcitedefaultmidpunct}
{\mcitedefaultendpunct}{\mcitedefaultseppunct}\relax
\EndOfBibitem
\bibitem{LHCb-PAPER-2019-001}
LHCb collaboration, R.~Aaij {\em et~al.},
  \ifthenelse{\boolean{articletitles}}{\emph{{Measurement of the mass
  difference between neutral charm-meson eigenstates}},
  }{}\href{https://doi.org/10.1103/PhysRevLett.122.231802}{Phys.\ Rev.\ Lett.\
  \textbf{122} (2019) 231802},
  \href{http://arxiv.org/abs/1903.03074}{{\normalfont\ttfamily
  arXiv:1903.03074}}\relax
\mciteBstWouldAddEndPuncttrue
\mciteSetBstMidEndSepPunct{\mcitedefaultmidpunct}
{\mcitedefaultendpunct}{\mcitedefaultseppunct}\relax
\EndOfBibitem
\bibitem{Sjostrand:2007gs}
T.~Sj\"{o}strand, S.~Mrenna, and P.~Skands,
  \ifthenelse{\boolean{articletitles}}{\emph{{A brief introduction to PYTHIA
  8.1}}, }{}\href{https://doi.org/10.1016/j.cpc.2008.01.036}{Comput.\ Phys.\
  Commun.\  \textbf{178} (2008) 852},
  \href{http://arxiv.org/abs/0710.3820}{{\normalfont\ttfamily
  arXiv:0710.3820}}\relax
\mciteBstWouldAddEndPuncttrue
\mciteSetBstMidEndSepPunct{\mcitedefaultmidpunct}
{\mcitedefaultendpunct}{\mcitedefaultseppunct}\relax
\EndOfBibitem
\bibitem{Sjostrand:2006za}
T.~Sj\"{o}strand, S.~Mrenna, and P.~Skands,
  \ifthenelse{\boolean{articletitles}}{\emph{{PYTHIA 6.4 physics and manual}},
  }{}\href{https://doi.org/10.1088/1126-6708/2006/05/026}{JHEP \textbf{05}
  (2006) 026}, \href{http://arxiv.org/abs/hep-ph/0603175}{{\normalfont\ttfamily
  arXiv:hep-ph/0603175}}\relax
\mciteBstWouldAddEndPuncttrue
\mciteSetBstMidEndSepPunct{\mcitedefaultmidpunct}
{\mcitedefaultendpunct}{\mcitedefaultseppunct}\relax
\EndOfBibitem
\bibitem{LHCb-PROC-2010-056}
I.~Belyaev {\em et~al.}, \ifthenelse{\boolean{articletitles}}{\emph{{Handling
  of the generation of primary events in Gauss, the LHCb simulation
  framework}}, }{}\href{https://doi.org/10.1088/1742-6596/331/3/032047}{J.\
  Phys.\ Conf.\ Ser.\  \textbf{331} (2011) 032047}\relax
\mciteBstWouldAddEndPuncttrue
\mciteSetBstMidEndSepPunct{\mcitedefaultmidpunct}
{\mcitedefaultendpunct}{\mcitedefaultseppunct}\relax
\EndOfBibitem
\bibitem{Lange:2001uf}
D.~J. Lange, \ifthenelse{\boolean{articletitles}}{\emph{{The EvtGen particle
  decay simulation package}},
  }{}\href{https://doi.org/10.1016/S0168-9002(01)00089-4}{Nucl.\ Instrum.\
  Meth.\  \textbf{A462} (2001) 152}\relax
\mciteBstWouldAddEndPuncttrue
\mciteSetBstMidEndSepPunct{\mcitedefaultmidpunct}
{\mcitedefaultendpunct}{\mcitedefaultseppunct}\relax
\EndOfBibitem
\bibitem{davidson2015photos}
N.~Davidson, T.~Przedzinski, and Z.~Was,
  \ifthenelse{\boolean{articletitles}}{\emph{{PHOTOS interface in C++:
  Technical and physics documentation}},
  }{}\href{https://doi.org/https://doi.org/10.1016/j.cpc.2015.09.013}{Comp.\
  Phys.\ Comm.\  \textbf{199} (2016) 86},
  \href{http://arxiv.org/abs/1011.0937}{{\normalfont\ttfamily
  arXiv:1011.0937}}\relax
\mciteBstWouldAddEndPuncttrue
\mciteSetBstMidEndSepPunct{\mcitedefaultmidpunct}
{\mcitedefaultendpunct}{\mcitedefaultseppunct}\relax
\EndOfBibitem
\bibitem{Allison:2006ve}
Geant4 collaboration, J.~Allison {\em et~al.},
  \ifthenelse{\boolean{articletitles}}{\emph{{Geant4 developments and
  applications}}, }{}\href{https://doi.org/10.1109/TNS.2006.869826}{IEEE
  Trans.\ Nucl.\ Sci.\  \textbf{53} (2006) 270}\relax
\mciteBstWouldAddEndPuncttrue
\mciteSetBstMidEndSepPunct{\mcitedefaultmidpunct}
{\mcitedefaultendpunct}{\mcitedefaultseppunct}\relax
\EndOfBibitem
\bibitem{Agostinelli:2002hh}
Geant4 collaboration, S.~Agostinelli {\em et~al.},
  \ifthenelse{\boolean{articletitles}}{\emph{{Geant4: A simulation toolkit}},
  }{}\href{https://doi.org/10.1016/S0168-9002(03)01368-8}{Nucl.\ Instrum.\
  Meth.\  \textbf{A506} (2003) 250}\relax
\mciteBstWouldAddEndPuncttrue
\mciteSetBstMidEndSepPunct{\mcitedefaultmidpunct}
{\mcitedefaultendpunct}{\mcitedefaultseppunct}\relax
\EndOfBibitem
\bibitem{LHCb-PROC-2011-006}
M.~Clemencic {\em et~al.}, \ifthenelse{\boolean{articletitles}}{\emph{{The
  \lhcb simulation application, Gauss: Design, evolution and experience}},
  }{}\href{https://doi.org/10.1088/1742-6596/331/3/032023}{J.\ Phys.\ Conf.\
  Ser.\  \textbf{331} (2011) 032023}\relax
\mciteBstWouldAddEndPuncttrue
\mciteSetBstMidEndSepPunct{\mcitedefaultmidpunct}
{\mcitedefaultendpunct}{\mcitedefaultseppunct}\relax
\EndOfBibitem
\bibitem{LHCb-DP-2018-004}
D.~M{\"u}ller, M.~Clemencic, G.~Corti, and M.~Gersabeck,
  \ifthenelse{\boolean{articletitles}}{\emph{{ReDecay: A novel approach to
  speed up the simulation at LHCb}},
  }{}\href{https://doi.org/10.1140/epjc/s10052-018-6469-6}{Eur.\ Phys.\ J.\
  \textbf{C78} (2018) 1009},
  \href{http://arxiv.org/abs/1810.10362}{{\normalfont\ttfamily
  arXiv:1810.10362}}\relax
\mciteBstWouldAddEndPuncttrue
\mciteSetBstMidEndSepPunct{\mcitedefaultmidpunct}
{\mcitedefaultendpunct}{\mcitedefaultseppunct}\relax
\EndOfBibitem
\bibitem{Breiman}
L.~Breiman, J.~H. Friedman, R.~A. Olshen, and C.~J. Stone, {\em Classification
  and regression trees}, Wadsworth international group, Belmont, California,
  USA, 1984\relax
\mciteBstWouldAddEndPuncttrue
\mciteSetBstMidEndSepPunct{\mcitedefaultmidpunct}
{\mcitedefaultendpunct}{\mcitedefaultseppunct}\relax
\EndOfBibitem
\bibitem{Roe}
B.~P. Roe {\em et~al.}, \ifthenelse{\boolean{articletitles}}{\emph{{Boosted
  decision trees as an alternative to artificial neural networks for particle
  identification}}, }{}\href{https://doi.org/10.1016/j.nima.2004.12.018}{Nucl.\
  Instrum.\ Meth.\  \textbf{A543} (2005) 577},
  \href{http://arxiv.org/abs/physics/0408124}{{\normalfont\ttfamily
  arXiv:physics/0408124}}\relax
\mciteBstWouldAddEndPuncttrue
\mciteSetBstMidEndSepPunct{\mcitedefaultmidpunct}
{\mcitedefaultendpunct}{\mcitedefaultseppunct}\relax
\EndOfBibitem
\bibitem{TMVA}
A.~{Hoecker} {\em et~al.}, \ifthenelse{\boolean{articletitles}}{\emph{{TMVA -
  Toolkit for Multivariate Data Analysis}}, }{}PoS \textbf{ACAT} (2007) 040,
  \href{http://arxiv.org/abs/physics/0703039}{{\normalfont\ttfamily
  arXiv:physics/0703039}}\relax
\mciteBstWouldAddEndPuncttrue
\mciteSetBstMidEndSepPunct{\mcitedefaultmidpunct}
{\mcitedefaultendpunct}{\mcitedefaultseppunct}\relax
\EndOfBibitem
\bibitem{AdaBoost}
Y.~Freund and R.~E. Schapire, \ifthenelse{\boolean{articletitles}}{\emph{A
  decision-theoretic generalization of on-line learning and an application to
  boosting}, }{}\href{https://doi.org/10.1006/jcss.1997.1504}{J.\ Comput.\
  Syst.\ Sci.\  \textbf{55} (1997) 119}\relax
\mciteBstWouldAddEndPuncttrue
\mciteSetBstMidEndSepPunct{\mcitedefaultmidpunct}
{\mcitedefaultendpunct}{\mcitedefaultseppunct}\relax
\EndOfBibitem
\bibitem{Blum1999BeatingTH}
A.~Blum, A.~Kalai, and J.~Langford,
  \ifthenelse{\boolean{articletitles}}{\emph{{Beating the hold-out: bounds for
  k-fold and progressive cross-validation}}, }{}COLT '99 Proceedings of the
  twelfth annual conference on Computational learning theory (1999) 203\relax
\mciteBstWouldAddEndPuncttrue
\mciteSetBstMidEndSepPunct{\mcitedefaultmidpunct}
{\mcitedefaultendpunct}{\mcitedefaultseppunct}\relax
\EndOfBibitem
\bibitem{LHCb-DP-2018-001}
R.~Aaij {\em et~al.}, \ifthenelse{\boolean{articletitles}}{\emph{{Selection and
  processing of calibration samples to measure the particle identification
  performance of the LHCb experiment in Run 2}},
  }{}\href{https://doi.org/10.1140/epjti/s40485-019-0050-z}{Eur.\ Phys.\ J.\
  Tech.\ Instr.\  \textbf{6} (2019) 1},
  \href{http://arxiv.org/abs/1803.00824}{{\normalfont\ttfamily
  arXiv:1803.00824}}\relax
\mciteBstWouldAddEndPuncttrue
\mciteSetBstMidEndSepPunct{\mcitedefaultmidpunct}
{\mcitedefaultendpunct}{\mcitedefaultseppunct}\relax
\EndOfBibitem
\bibitem{supplemental}
See supplemental material for the efficiency map of \Lcpmm decays, the
  dimuon-mass resolution, as well as a table of the systematic uncertainties
  and the correlation matrix for the measured asymmetries.\relax
\mciteBstWouldAddEndPunctfalse
\mciteSetBstMidEndSepPunct{\mcitedefaultmidpunct}
{}{\mcitedefaultseppunct}\relax
\EndOfBibitem
\bibitem{Johnson:1949zj}
N.~L. Johnson, \ifthenelse{\boolean{articletitles}}{\emph{{Systems of frequency
  curves generated by methods of translation}},
  }{}\href{https://doi.org/10.1093/biomet/36.1-2.149}{Biometrika \textbf{36}
  (1949) 149}\relax
\mciteBstWouldAddEndPuncttrue
\mciteSetBstMidEndSepPunct{\mcitedefaultmidpunct}
{\mcitedefaultendpunct}{\mcitedefaultseppunct}\relax
\EndOfBibitem
\bibitem{Skwarnicki:1986xj}
T.~Skwarnicki, {\em {A study of the radiative cascade transitions between the
  Upsilon-prime and Upsilon resonances}}, PhD thesis, Institute of Nuclear
  Physics, Krakow, 1986,
  {\href{http://inspirehep.net/record/230779/}{DESY-F31-86-02}}\relax
\mciteBstWouldAddEndPuncttrue
\mciteSetBstMidEndSepPunct{\mcitedefaultmidpunct}
{\mcitedefaultendpunct}{\mcitedefaultseppunct}\relax
\EndOfBibitem
\bibitem{efron:1979}
B.~Efron, \ifthenelse{\boolean{articletitles}}{\emph{Bootstrap methods: Another
  look at the jackknife},
  }{}\href{https://doi.org/10.1214/aos/1176344552}{Ann.\ Statist.\  \textbf{7}
  (1979) 1}\relax
\mciteBstWouldAddEndPuncttrue
\mciteSetBstMidEndSepPunct{\mcitedefaultmidpunct}
{\mcitedefaultendpunct}{\mcitedefaultseppunct}\relax
\EndOfBibitem
\bibitem{PhysRevLett.42.13}
A.~Gsponer {\em et~al.}, \ifthenelse{\boolean{articletitles}}{\emph{Precise
  coherent ${K}_{S}$ regeneration amplitudes for c, al, cu, sn, and pb nuclei
  from 20 to 140 gev/c and their interpretation},
  }{}\href{https://doi.org/10.1103/PhysRevLett.42.13}{Phys.\ Rev.\ Lett.\
  \textbf{42} (1979) 13}\relax
\mciteBstWouldAddEndPuncttrue
\mciteSetBstMidEndSepPunct{\mcitedefaultmidpunct}
{\mcitedefaultendpunct}{\mcitedefaultseppunct}\relax
\EndOfBibitem
\bibitem{PhysRevLett.75.402}
R.~A. Briere and B.~Winstein,
  \ifthenelse{\boolean{articletitles}}{\emph{Determining the phase of a strong
  scattering amplitude from its momentum dependence to better than
  1\ifmmode^\circ\else\textdegree\fi{}: The example of kaon regeneration},
  }{}\href{https://doi.org/10.1103/PhysRevLett.75.402}{Phys.\ Rev.\ Lett.\
  \textbf{75} (1995) 402}\relax
\mciteBstWouldAddEndPuncttrue
\mciteSetBstMidEndSepPunct{\mcitedefaultmidpunct}
{\mcitedefaultendpunct}{\mcitedefaultseppunct}\relax
\EndOfBibitem
\bibitem{LHCb-PAPER-2014-013}
LHCb collaboration, R.~Aaij {\em et~al.},
  \ifthenelse{\boolean{articletitles}}{\emph{{Measurement of \CP asymmetry in
  \mbox{\decay{\Dz}{\Km\Kp}} and \mbox{\decay{\Dz}{\pim\pip}} decays}},
  }{}\href{https://doi.org/10.1007/JHEP07(2014)041}{JHEP \textbf{07} (2014)
  041}, \href{http://arxiv.org/abs/1405.2797}{{\normalfont\ttfamily
  arXiv:1405.2797}}\relax
\mciteBstWouldAddEndPuncttrue
\mciteSetBstMidEndSepPunct{\mcitedefaultmidpunct}
{\mcitedefaultendpunct}{\mcitedefaultseppunct}\relax
\EndOfBibitem
\bibitem{LHCb-PAPER-2022-024}
LHCb collaboration, R.~Aaij {\em et~al.},
  \ifthenelse{\boolean{articletitles}}{\emph{{Measurement of the
  time-integrated \CP asymmetry in $\Dz \to \Km \Kp$ decays}},
  }{}\href{https://doi.org/10.1103/PhysRevLett.131.091802}{Phys.\ Rev.\ Lett.\
  \textbf{131} (2023) 091802},
  \href{http://arxiv.org/abs/2209.03179}{{\normalfont\ttfamily
  arXiv:2209.03179}}\relax
\mciteBstWouldAddEndPuncttrue
\mciteSetBstMidEndSepPunct{\mcitedefaultmidpunct}
{\mcitedefaultendpunct}{\mcitedefaultseppunct}\relax
\EndOfBibitem
\bibitem{LYONS1988110}
L.~Lyons, D.~Gibaut, and P.~Clifford,
  \ifthenelse{\boolean{articletitles}}{\emph{How to combine correlated
  estimates of a single physical quantity},
  }{}\href{https://doi.org/https://doi.org/10.1016/0168-9002(88)90018-6}{Nucl.\
  Instrum.\ Meth.\  \textbf{A270} (1988) 110}\relax
\mciteBstWouldAddEndPuncttrue
\mciteSetBstMidEndSepPunct{\mcitedefaultmidpunct}
{\mcitedefaultendpunct}{\mcitedefaultseppunct}\relax
\EndOfBibitem
\bibitem{LHCb-DP-2022-002}
LHCb collaboration, R.~Aaij {\em et~al.},
  \ifthenelse{\boolean{articletitles}}{\emph{{The LHCb Upgrade I}},
  }{}\href{https://doi.org/10.1088/1748-0221/19/05/P05065}{{JINST} \textbf{19}
  (2024) P05065}, \href{http://arxiv.org/abs/2305.10515}{{\normalfont\ttfamily
  arXiv:2305.10515}}\relax
\mciteBstWouldAddEndPuncttrue
\mciteSetBstMidEndSepPunct{\mcitedefaultmidpunct}
{\mcitedefaultendpunct}{\mcitedefaultseppunct}\relax
\EndOfBibitem
\bibitem{data}
https://cds.cern.ch/record/2923688.
\newblock Data associated to the plots in this publication as well as in
  supplementary materials\relax
\mciteBstWouldAddEndPuncttrue
\mciteSetBstMidEndSepPunct{\mcitedefaultmidpunct}
{\mcitedefaultendpunct}{\mcitedefaultseppunct}\relax
\EndOfBibitem
\end{mcitethebibliography}
\ifx\mcitethebibliography\mciteundefinedmacro
\PackageError{LHCb.bst}{mciteplus.sty has not been loaded}
{This bibstyle requires the use of the mciteplus package.}\fi
\providecommand{\href}[2]{#2}

\clearpage
\newpage
%
%
%
\centerline
{\large\bf LHCb collaboration}
\begin
{flushleft}
\small
R.~Aaij$^{38}$\lhcborcid{0000-0003-0533-1952},
A.S.W.~Abdelmotteleb$^{57}$\lhcborcid{0000-0001-7905-0542},
C.~Abellan~Beteta$^{51}$\lhcborcid{0009-0009-0869-6798},
F.~Abudin{\'e}n$^{57}$\lhcborcid{0000-0002-6737-3528},
T.~Ackernley$^{61}$\lhcborcid{0000-0002-5951-3498},
A. A. ~Adefisoye$^{69}$\lhcborcid{0000-0003-2448-1550},
B.~Adeva$^{47}$\lhcborcid{0000-0001-9756-3712},
M.~Adinolfi$^{55}$\lhcborcid{0000-0002-1326-1264},
P.~Adlarson$^{83}$\lhcborcid{0000-0001-6280-3851},
C.~Agapopoulou$^{14}$\lhcborcid{0000-0002-2368-0147}
% ...

{\footnotesize \it

% ...
$^{14}$Universit{\'e} Paris-Saclay, CNRS/IN2P3, IJCLab, Orsay, France\\
% ...
$^{38}$Nikhef National Institute for Subatomic Physics, Amsterdam, Netherlands\\
% ...
$^{47}$Instituto Galego de F{\'\i}sica de Altas Enerx{\'\i}as (IGFAE), Universidade de Santiago de Compostela, Santiago de Compostela, Spain\\
% ...
$^{51}$Physik-Institut, Universit{\"a}t Z{\"u}rich, Z{\"u}rich, Switzerland\\
% ...
$^{55}$H.H. Wills Physics Laboratory, University of Bristol, Bristol, United Kingdom\\
% ...
$^{57}$Department of Physics, University of Warwick, Coventry, United Kingdom\\
% ...
$^{60}$School of Physics and Astronomy, University of Glasgow, Glasgow, United Kingdom\\
$^{61}$Oliver Lodge Laboratory, University of Liverpool, Liverpool, United Kingdom\\
% ...
$^{69}$Syracuse University, Syracuse, NY, United States\\
% ...
$^{83}$Department of Physics and Astronomy, Uppsala University, Uppsala, Sweden, associated to $^{60}$\\
% ...
}
\end{flushleft} 
 
\end{document}